\documentclass[preprint,11pt]{JHEP3} 

\JHEPspecialurl{http://jhep.sissa.it/JOURNAL/JHEP3.tar.gz}

\usepackage{epsfig,multicol,amsmath}

\def\beq{\begin{equation}}
\def\eeq{\end{equation}}
\def\bea{\begin{eqnarray}}
\def\eea{\end{eqnarray}}
\def\eq#1{{Eq.~(\ref{#1})}}
\def\fig#1{{Fig.~\ref{#1}}}

\newcommand{\Lb}{\left(}
\newcommand{\Rb}{\right)}
\parskip=\itemsep               

\renewcommand{\thefootnote}{\fnsymbol{footnote}}

\newcommand{\al}{\alpha}

\makeatletter \@addtoreset{equation}{section}
\makeatother \makeatletter \@addtoreset{equation}{section}

\def\thefootnote{\fnsymbol{footnote}}

\title{\Large \bf Modified Kolmogorov Wave Turbulence in QCD matched onto ``Bottom-up'' Thermalization
}
{\author{\large
Vladimir Khachatryan$^{\,a,b}$\,\thanks{Email: khavladi@yerphi.am.} \\
a) Theory Division, Yerevan Physics Institute, Yerevan, 375036, Armenia \\
b) HEP Department, School of Physics and Astronomy\\
\,\,\,\,\,\,\,Raymond and Beverly Sackler Faculty of Exact Science\\
\,\,\,\,\,\,\,Tel Aviv University, Tel Aviv, 69978, Israel \\
}
}

\abstract{We investigate modification of Kolmogorov wave turbulence in QCD calculating gluon spectra as functions of time in
the presence of a low energy source which feeds in energy density in the infrared region at a time-dependent rate. Then
considering the picture of saturation constraints as has been constructed in the ``bottom-up'' thermalization approach we
revisit that picture for RHIC center-mass energy, \,$W = 130\,GeV$,\, and also extend it to LHC center-mass energy,
\,$W = 5500 GeV$,\, thus for two cases having an opportunity to calculate the equilibration time, \,$\tau_{eq|therm}$,\,
of the gluon system produced in a central heavy ion collision at mid-rapidity region. Thereby, at RHIC and LHC energies we
can match the equilibration time, obtained from the late stage gluon spectrum of the modified Kolmogorov wave turbulence,
onto that of the ``bottom-up'' thermalization and other evolutional approaches as well. In addition, from the revised
``bottom-up'' approach we find the gluon liberation coefficient to be on the average, \,$\varepsilon\simeq 0.81 - 1.06$\,
at RHIC and \,$\varepsilon \simeq 0.50 - 0.56$\, at LHC. We also present other phenomenological estimates of \,$\tau_{therm}$\,
which, at QCD realistic couplings, yield \,$0.45\,fm - 0.65\,fm \leq \tau_{therm} \leq 0.97\,fm - 2.72\,fm$\, at RHIC and
\,$0.31\,fm - 0.40\,fm \leq \tau_{therm} \leq 0.86\,fm - 2.04\,fm$\, at LHC. We show that the second upper-bounds of
\,$\tau_{therm}$\, in both cases are due to the late stage gluon spectrum of the original Kolmogorov wave turbulence in QCD,
previously deduced with a low energy source which feeds in energy density at a constant rate. On the other hand, the
lower-bounds and first upper-bounds of \,$\tau_{therm}$\, are due to the late stage gluon spectrum of the modified QCD
wave turbulence, deduced here at the specific time-dependent rate. In the latter case, at certain conditions, taking also
into account both very small and realistic couplings we give estimates \,$-$\, \,$0.65\,fm \leq \tau_{therm} \leq 1.29\,fm$\,
at RHIC and \,$0.52\,fm \leq \tau_{therm} \leq 1.16\,fm$\, at LHC, as well as at realistic couplings we find
\,$0.53\,<\,\tau_{therm}\,<\,0.7\,fm$\, at RHIC and \,$0.41\,<\,\tau_{therm}\,<\,0.65\,fm$\, at LHC.}

\keywords{Kolmogorov Wave Turbulence, Kolmogorov gluon spectra, Heavy Ion Collisions, Saturation, ``Bottom-up''
thermalization, Equilibration}

\begin{document}

\def\thefootnote{\arabic{footnote}}
\section{Introduction}
\label{sec:Inr}
We attempt to continue the investigations which are based on ideas put forward in \cite{Mu99,BaMu,BMSS,MuShoWo05,MuShoWo06}
by Mueller and his colleagues. These articles are devoted to studies of early stages of the gluon system produced in high
energy collisions of heavy ions ($AA$) where the main attention is directed to solving the key question of physics of heavy
ion collisions: the thermalization of the produced gluon system. In this introductory section we wish to remind the
basic results of these and other related studies noting that our ansatz for further proceeding is based on Ref.\,\cite{MuShoWo06}.

Based on the gluon saturation scenario \cite{GLR,Mu90}, originally, studies of the equilibration picture were performed
in \cite{Mu99} where the initial conditions of the produced gluon system are taken from the McLerran-Venugopalan model
assuming that all saturated gluons, those having transverse momentum at/or below saturation momentum, $Q_{s}$, in the heavy
ion light cone wavefunction, are freed in the high energy collision while the gluons beyond $Q_{s}$ are not freed
\cite{McVe,JaKo,Ko96,KoMu}. Afterwards, the whole discussion was reformulated in a much more general way in terms of the
Boltzmann equation \cite{LiPi} with a collision term taken from the elastic gluon-gluon scattering in the one gluon exchange
approximation. Here the Boltzmann equation without particle production was considered: the $2\leftrightarrow2$ process
wherein there is a cancellation between gain and loss terms in the elastic collision integral. For this process the time
of kinetic equilibration during of which the initial gluon distribution changes significantly is of the order of
\,$\Lb \exp(const/\sqrt{\al})\,Q_{s}^{-1} \Rb$\, which is obtained taking into account a lot of the gluon-gluon elastic
scatterings at small angles.

With inclusion of the particle production into the Boltzmann equation  the gluon system does seem to approach the kinetic
equilibration during a time of the order of \,$\Lb \sim\,\al^{-13/5}Q_{s}^{-1} \Rb$\, as obtained in \cite{BaMu}. The
thermalization occurs in the limit \,$Q_{s} \gg \Lambda_{QCD}$\, which corresponds to very large nuclei or very high
collision energy. The interaction process is described by the ``bottom-up'' thermalization scenario in $AA$
collisions which is an attempt to study the different stages of evolution of the produced system of the gluons up to
the equilibration stage. Furthermore, this scenario provides to some extent agreement with experimental data in
respect to hadron multiplicities \cite{BMSS}. However, it was later realized that collective effects in the form of magnetic
plasma instabilities, known as Weibel or filamentary instabilities, necessarily play a role in the initial stage \footnote{
In very initial stage the particle momentum distribution is highly asymmetric and the source of the instabilities is a
collection of these hard particles with such an asymmetrical momentum distribution.} of the ``bottom-up'' equilibration
\cite{ArLeMo03} (see \cite{Weibel} and \cite{Mrow} for early discussions as well). Due to the plasma instabilities present
in the dense gluon system produced immediately after the collision, the amount of energy transformation, from initially
produced hard gluons into softer gluons radiated afterwards by the hard ones, is increased by which one can hope for a
rapid thermalization scenario \footnote{In more general terms the instabilities initially grow exponentially quickly,
expressed by creation of transverse chromomagnetic-electric fields at short times \cite{Strick} which could speed up local
isotropization and thermalization of the initially non-equilibrium plasma by scattering the plasma particles into random
directions.}. Nonetheless, it was shown \cite{ArLe} that the full equilibration time in the presence of the instabilities
is not much shorter relative to that of the ``bottom-up'' approach. But the instabilities cannot lead directly to
equilibration since they would give an equilibration time parametrically on the order of \,$Q_{s}\tau\,\sim\,1$.\, For a
more detailed insights in this current problem scaling solutions were obtained \cite{MuShoWo05} for pre-equilibrium
evolution between the instabilities of the initial stage and the final equilibration. These solutions, depending on one
single parameter, match onto the intermediate stage and/or the late stage of the evolution of the gluon system given by
the ``bottom-up'' thermalization which is otherwise called as modified ``bottom-up'' thermalization (hereafter referred to
as {\em m}``bottom-up''). Meanwhile, the problems to follow analytically how QCD gluonic system evolves towards the
equilibration stage in the presence of the instabilities have not been completely overcome so far (see also \cite{Bo}).
Numerical simulations also seem to indicate that the instabilities are effective at early times. In this regard we hope for
further continuous progress of serious numerical studies which can be found in
Refs.\,\cite{Strick,Bod,ArLea,ArMo,DuNa,RoSt,ReRoSt,RoVe}.

Afterwards, the Kolmogorov wave turbulence in QCD (or QCD wave turbulence) was explored in \cite{MuShoWo06} for
further better understanding of the instabilities problem in the early stages of the evolution after $AA$ collision,
performing calculations for finding time dependences of gluonic spectra, $f_{k}(t)$, in the presence of a low (infrared)
energy source which supplies energy at a constant rate to high energetic gluons. In general, the wave turbulence problem
\cite{ZLF,MiT} (with the low energy source) discussed by Zakharov, L'vov, Falkovich (ZLF) has some generalities with the
problem of the instabilities. However, the ZLF and QCD turbulences are somewhat different from each other. In the first
case the waves, or particles, interact with each other locally in momentum, e.g., as in a $\Phi^4$ theory. But in the
second case in QCD the soft and hard gluons have strong interactions along with a lower cutoff in frequency (the so-called
plasma frequency which is absent in the $\Phi^4$-type theories) or, in other words, the interactions are non-local in
momentum modes.

In this paper we attempt to modify the Kolmogorov wave turbulence in QCD. As in \cite{MuShoWo06} the used dynamics stems
from the Boltzmann equation with a collision term consisting of \,$2 \leftrightarrow 2$\, and \,$2 \leftrightarrow 3$\, gluon
processes. The absence of longrange coherent fields is similarly assumed. Afterwards, the received equilibration time can be
matched onto that from various evolutional approaches.

The paper is organized as follows. In the next section we discuss the time stages of the  ``bottom-up'' thermalization and
QCD wave turbulence.
In the third section we find an early, an intermediate and a late time analytic forms of the gluon spectra in the
presence of the low energy source which feeds in energy density at the time-dependent rate
\,$\dot{\epsilon}_{0}\,=\,\frac{m_{0}^{5}}{\al}\,\frac{2}{1 + e^{\gamma(\tau/\tau_{0})}}$,\, where \,$\al$\,
is the QCD coupling constant supposed to be small, $\tau$ is the proper time of the central collision region,
\,$\tau_{0} = 1\,fm$\, and \,$\gamma$\, is a parametric constant the meaning of which will be explained afterwards
\footnote{The case using \,$\,\dot{\epsilon}_{0}\,=\,m_{0}^{5}/\al\,=\,const\,$\, for the investigation of the early,
intermediate and late time forms of the gluonic spectra is already done in \cite{MuShoWo06} with $m_{0}$, being a single
dimensionful parameter, and assuming $\al$ to be small.}. For the calculations of the gluon spectra we will follow along
the lines of Ref.\,\cite{MuShoWo06}, again supposing that the energy is incoming into our system uniformly in space in the
form of the gluons momentum distribution of which are spherically symmetric. These gluons are distributed in phase space
uniformly in a range \,$m\,<\,\omega\,<\,\bar{m}$\, with $m$ to be proportional to the gluon plasma frequency (which
defines the soft scale), and $\bar{m}$ on the order of $m$. In the fourth section we calculate the gluon liberation
coefficient, $\varepsilon$ (in a limited range), at RHIC and LHC, using the procedure for finding the constraints
of the saturation as was done in \cite{BMSS}. Making use of $\varepsilon$ we find a constant factor \,$\varepsilon_{eq}$\,
in the equilibration time formula \,$\Lb \tau_{eq}\,=\,\varepsilon_{eq}\,\al^{-13/5}Q_{s}^{-1} \Rb$\, of the
{\em m}``bottom-up''/``bottom-up'' thermalization. Thus at RHIC and LHC energies we match the equilibration time, obtained
from the late stage gluon spectrum of the modified QCD wave turbulence, onto that of the {\em m}``bottom-up''/``bottom-up''
thermalization. Then one can see that depending on the value of $\gamma$ and $\al$ the matching can occur or not. Finally,
in the section fifth we discuss and summarize our results.

\section{The time stages of the ``bottom-up'' thermalization and QCD wave turbulence}
\label{sec:Botup}
\subsection{The time stages of the ``bottom-up'' thermalization}
The conventional argument in favor of the thermalization is that at higher collision energy more gluons are freed at a time
around $1/Q_{s}$. In the original ``bottom-up'' picture \cite{BaMu} these gluons (the so-called hard gluons), having
momentum at or near the saturation scale of the colliding nuclei,  produced after the nuclear impact, lose energy
by radiating soft gluons. Hereon, the number of these soft gluons becomes significantly large such that they equilibrate
amongst themselves forming a thermal bath which continues to draw energy from the initially produced hard gluons. The full
equilibration is achieved when the hard gluons have lost all their energy. The whole process is divided into three
distinct stages:
\begin{itemize}
\item[a)] the early stage \,$1\,<\,Q_{s}\tau\,<\,\al^{-3/2}$\, (the hard gluons dominate);
\item[b)] the intermediate stage \,$\al^{-3/2}\,<\,Q_{s}\tau\,<\,\al^{-5/2}$\, (the hard gluons still dominate in
number but the occupation number $f_{h}\,<\,1$);
\item[c)] the late stage \,$\al^{-5/2}\,<\,Q_{s}\tau\,<\,\al^{-13/5}$\, (the soft gluons dominate over the hard
ones and the system reaches thermal equilibrium).
\end{itemize}
But as observed in \cite{ArLeMo03} the early stage of the ``bottom-up'' is not correct because here instead of
having a screening mass (Debye mass, $m_{D}$) the system of the produced hard gluons, having very asymmetrical momentum
distribution, leads to a mass characterizing the instabilities. Besides, it is not immediately clear whether at
intermediate stage the ``bottom-up'' is self-consistent. Meanwhile, its late stage part should be consistent since
there the Debye mass is determined from the thermalized soft gluons.

However, in the {\em m}``bottom-up'' thermalization for gluons produced at time $\tau$ \footnote{Throughout paper we
will be interested only in the central rapidity region of central collisions. In this region one can assume boost invariance
whereby all physical quantities depend only on the proper time $\tau$.} there are scaling solutions \cite{MuShoWo05} which
allow us interpolate between the instabilities and equilibration. This one-parameter family of the scaling solutions,
parametrized by a positive number $\delta$, is the following:
$$
N_{s}\,\sim\,\frac{Q_{s}^3}{\al(Q_{s}\tau)^{4/3-\delta}}\,,\,\,\,\,\,\,\,\,\,\,\,\,\,\,\,k_{s}\,\,\sim\,\,
\frac{Q_{s}}{(Q_{s}\tau)^{1/3-2\delta/5}}\,,
$$
\beq \label{ScSu}
\,\,\,\,\,
f_{s}\,\sim\,\frac{1}{\al(Q_{s}\tau)^{1/3+\delta/5}}\,,\,\,\,\,\,\,\,\,\,\,\,m_{D}\,\sim\,
\frac{Q_{s}}{(Q_{s}\tau)^{1/2-3\delta/10}}\,
\eeq
where $N_{s}$ is the number density of the soft gluons, $k_{s}$ - the soft gluon momentum, $f_{s}$ - the soft gluon
occupation number and the symbol \,$\sim$\, means that we are still unable to evaluate the total constant factor of
each of these expressions.
The proposed solutions become identical to the intermediate stage of the ``bottom-up'' picture in the range
\,$0 \leq \delta \leq 1/3$.\, And for the point \,$\delta = 1/3$\, we have the case that the transition from the scaling
solutions to the ``bottom-up'' solution is to the beginning of the late stage of the ``bottom-up'', starting at
\,$Q_{s}\tau\,\sim\,\al^{-5/2}$.\, When \,$\delta > 1/3$\, the coincidence with the ``bottom-up'' occurs at the final
time $Q_{s}\tau\,\sim\,\al^{-13/5}$.

\subsection{The time stages of QCD wave turbulence}
Now we mention the main conclusions from the recent investigation of QCD wave turbulence \cite{MuShoWo06} in view of the
corresponding time stages obtained from the calculations of the gluon spectra as functions of time. Here there are also
three distinct stages:
\begin{itemize}
\item[a*)] the early stage \,$1\,\ll\,m_{0}\tau\,<\,\al^{-7/5}$\, (the system is far from the thermal equilibrium and the
hard gluons dominate);
\item[b*)] the intermediate stage \,$\al^{-7/5}\,<\,m_{0}\tau\,<\,\al^{-9/5}$\, (the gluons having momentum much
greater than $m$ are in the thermal equilibrium but the occupation numbers do not coincide with the thermal curve in
the domain $m\,<\,\omega\,<\,\bar{m}$);
\item[c*)] the late stage \,$m_{0}\tau\,>\,\al^{-9/5}$\, (the system is very close to the thermal equilibrium where
the incoming energy is transferred from the soft scale $m$ to the hard scale given by the temperature $T$ by direct
absorption of the soft gluons by the hard gluons in an inelastic $3\rightarrow2$ process and, parametrically of the same
order, by an elastic scattering of the soft gluons on the hard ones with the soft gluons losing energy to them).
\end{itemize}
We point out that the ``bottom-up'' thermalization is also based on the observation that both inelastic and elastic processes
are equally important for the thermalization \cite{Wo96}. Thus we see that these two approaches have some similarities
albeit they are opposite in nature since in QCD wave turbulence we have the energy flow from the soft to hard modes while
in the ``bottom-up'' thermalization we have the energy transformation from the hard into soft modes.

\section{The spectrum of the gluons as a function of time and flow weakening parameter}
\label{sec:Spectrum}
Suppose that at time \,$\tau\,=\,0$\, we turn on the low energy source of the soft gluons which feeds in the energy density
at the rate \,$\dot{\epsilon}_{0}\,=\,\frac{m_{0}^{5}}{\al}\,\frac{2}{1 + e^{\gamma(\tau/\tau_{0})}}$.\, Or in other words
we conjecture that the energy amount deposited in the soft gauge sector is not a constant in time \cite{MuShoWo06} but
instead decreases exponentially with a damping factor, $\gamma$.
As noted in the introduction the energy enters into our system isotropically in low momentum modes, nevertheless, as the
density of the gluons increases in time the source is modified so that the incoming gluons always have energy just above $m$.
Why we take the time-dependent rate for feeding in the energy density will be clear later on when we match the equilibration
 time from the late stage spectrum onto the equilibration time of the ``bottom-up'' thermalization.

\subsection{The spectrum for\,\, $m_{0} \Psi(\gamma,\tau) > \al^{-9/5}$}
It is more straightforward to start from the calculation of the late time spectrum and in order to carry out it, first, we
keep the dimensions true, namely
\beq \label{Sp1}
\dot{\epsilon}(\tau)\,=\,\frac{m_{0}^{5}}{\al}\,\frac{2}{1 + e^{\gamma(\tau/\tau_{0})}}
\eeq
where $\gamma$ takes the numbers from $0$ up to $6$: integers and non-integers. The parameter $\gamma$ can also get higher
values but we will be restricted up to the number $6$. It describes the degree of the energy flow weakening from the soft
to hard scales \,\footnote{In our consideration the energy flow of the incoming gluons is somewhat similar to the so-called
``avalanche'' which is observed in the field isotropization driven by the plasma instabilities \cite{DuNa}.}.

\hskip -0.7truecm
The energy conservation gives
\beq \label{Sp2}
\epsilon(\tau)\,=\,\frac{m_{0}^{5}}{\al}\,2\Lb \tau - \frac{\tau_{0}}{\gamma}\ln\!\!\Lb
\frac{1 + e^{\gamma(\tau/\tau_{0})}}{2} \Rb \Rb
\,=\,g_{E}T^{4}
\eeq

\FIGURE[ht]{
\centerline{\epsfig{file=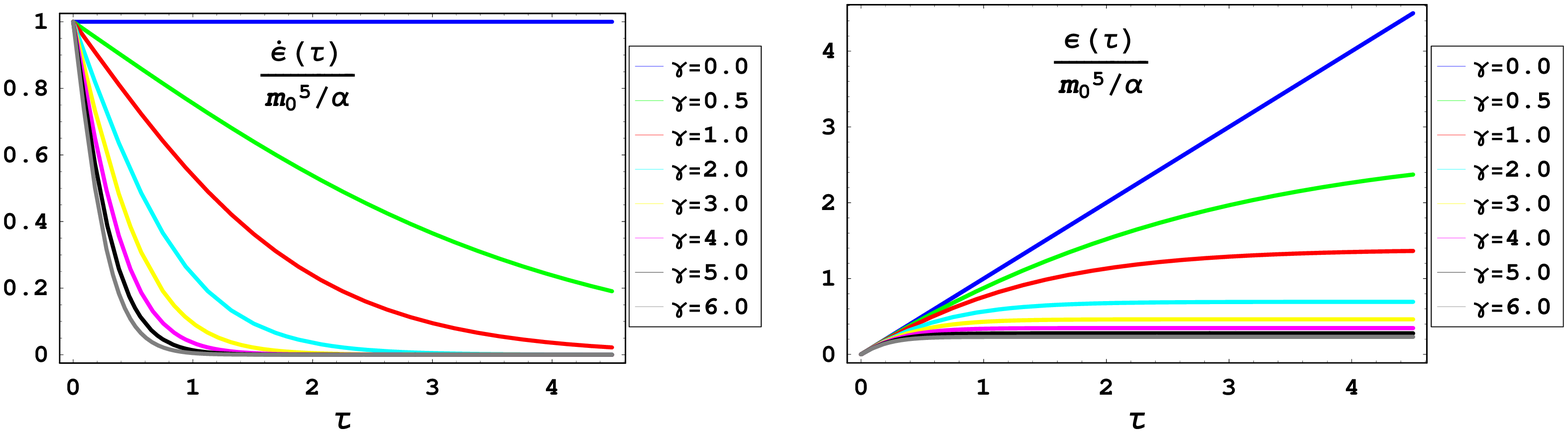,width=185mm,height=50mm}}
\caption{The rate feeding in the energy density (left panel) and the energy density (right panel) as a function
of time.}
\label{rate_energy} }

\hskip -0.7truecm
with \,$g_{E}\,=\,2(N_{c}^{2}-1)\frac{\pi^{2}}{30}$\, (see \fig{rate_energy}). Thence
\beq \label{Sp3}
T(\tau)\,=\,m_{0}\left[\frac{2\,m_{0}}{g_{E}\,\al} \Lb \tau - \frac{\tau_{0}}{\gamma}\ln\!\!
\Lb \frac{1 + e^{\gamma(\tau/\tau_{0})}}{2} \Rb \Rb \right]^{1/4}\,.
\eeq
We notice that in the limit of $\gamma\,\rightarrow\,0$ we have the case of the study \cite{MuShoWo06}.
The soft $m$ and hard $T$ scales are related to each other through\, $m \sim \omega_{P}^{}$\, where the gluon plasma frequency\,\,
$
\omega_{P}^{} = m_{\infty}\sqrt{1 + \eta\sqrt{4\pi\al N_{c}}}
$\,\,
with\,\,
$
m_{\infty} = \sqrt{(4\pi/9)N_{c}}\,\sqrt{\al}\,T
$\,\, \cite{Schu}.
$m_{\infty}$ is also linked to the Debye mass by \,$m_{\infty}\,=\,m_{D}/\sqrt{2}$\, and the coefficient $\eta$ is equal to
 \,$-0.18$. Nonetheless, for realistic small $\al$ the plasma frequency can be approximated as follows:
\beq \label{Sp4}
\omega_{P}^{}\,=\,m_{\infty}\sqrt{1 + \eta\sqrt{4\pi\al N_{c}}}\,\approx\,\sqrt{\frac{4\pi}{9}N_{c}}\,
\sqrt{\frac{\al}{2.7}}\,\,T\,\,\,\,\,\Rightarrow\,\,\,\,\,m\,\sim\,\sqrt{\al}\,T
\eeq
(at very small $\al$'s the denominator $2.7$ in the square root can be replaced by unity).

Consider the elastic scattering of the soft gluons, having momenta $q_{1}$ and $q_{2}$, on the hard gluons, having
momenta $p_{1}$ and $p_{2}$ which are of the order of \,$T$. \,In other words, by this process (see \fig{elastic}) the
soft gluons directly transfer energy to the hard ones.

\FIGURE{
\centerline{\epsfig{file=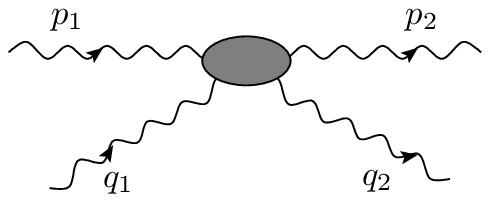,width=62mm,height=27mm}}
\caption{The elastic scattering of the soft and hard gluons. The figure is from \cite{MuShoWo06}.}
\label{elastic} }

\vskip 0.0truecm %
\hskip -0.7truecm %
The elastic rate is
{\setlength\arraycolsep{2pt}
\bea \label{Sp5}
\dot{\epsilon}^{el} & = & \frac{[2(N_{c}^{2}-1)]^{2}}{(2\pi)^{12}}\int_{R}\frac{d^{3}p_{1}}{2E_{1}}\frac{d^{3}p_{2}}{2E_{2}}
\frac{d^{3}q_{1}}{2\omega_{1}}\frac{d^{3}q_{2}}{2\omega_{2}}\,(2\pi)^{4}\delta^{4}(p_{1} + q_{1} - p_{2} - q_{2})\,
|M|^{2}\,\times
\nonumber\\
& \times &
\,[f_{p_{1}}f_{q_{1}}(1 + f_{p_{2}})(1 + f_{q_{2}}) - f_{p_{2}}f_{q_{2}}(1 + f_{p_{1}})(1 + f_{q_{1}})]\,(\omega_{1} -
\omega_{2})
\eea}
where $|M|^{2}$ is the gluon-gluon elastic scattering amplitude
\beq \label{Sp6}
|M|^{2}\,=\,\frac{64\pi^{4}}{N_{c}^{2} - 1}\Lb \frac{\al N_{c}}{\pi} \Rb^{2}\Lb 3 - \frac{ut}{s^{2}} - \frac{us}{t^{2}} -
 \frac{ts}{u^{2}} \Rb
\eeq
with
\beq \label{Sp7}
s\,=\,(p_{1}+q_{1})^{2}\,,\,\,\,\,\,t\,=\,(q_{1} - q_{2})^{2}\,,\,\,\,\,\,u\,=\,(p_{1} - q_{1})^{2}\,.
\eeq
The symbol $R$ under the integral sign in \eq{Sp5} restricts $\omega_{1},\omega_{2}$ to be less than $E_{1},E_{2}$.
The magnitudes $\omega_{1}$, $\omega_{2}$, $E_{1}$ and $E_{2}$ are the energies of the soft and hard gluons, respectively.

Let us stress that if the system were in exact equilibrium there would not be flow of the energy and particle number
and in that case the spectrum would be given as follows:
\beq \label{Sp8}
f_{q_{1},q_{2}}\,=\,\frac{1}{e^{\frac{\omega_{1},\omega_{2}}{T}} - 1}\,,\,\,\,\,\,\,\,\,\,\,\,
f_{p_{1},p_{2}}\,=\,\frac{1}{e^{\frac{E_{1},E_{2}}{T}}-1}\,,
\eeq
wherefrom $f_{q}$ can be approximated as
\beq \label{Sp9}
f_{q}\,\simeq\,\frac{T}{\omega}
\eeq
when $\omega/T \ll 1$.
Using \eq{Sp8} for the hard particles we also have
\beq \label{Sp10}
1 + f_{p}\,=\,f_{p}\,e^{E/T}
\eeq
which allows us to write the bracket in \eq{Sp5} as follows (with the use of the energy conservation)
\beq \label{Sp11}
[\,\,]\,=\,e^{E_{1}/T}f_{p_{1}}f_{p_{2}}f_{q_{1}}f_{q_{2}}\,[(\omega_{1}-\omega_{2})/T + 1/f_{q_{2}} - 1/f_{q_{1}}]\,.
\eeq
The hard momenta are designated as \,$p_{i}\,=\,(E_{i},\vec{p}_{i})$\, with \,$E_{i}\,\approx\,|\vec{p}_{i}|$\, while the
soft momenta read \,$q_{j}\,=\,(\omega_{j},\vec{q}_{j})$\, with \,$\omega_{j}^{2}\,=\,m^{2} + \vec{q}_{j}^{\,2}\,\approx\,m^{2}$.
\,The dominant contribution to $|M|^{2}$ comes from the small-$t$ region where \,$u\,\approx\,-s$,\, so that \eq{Sp6} switches
over to the form
\beq \label{Sp12}
|M|^{2}\,\sim\,\al^{2}\frac{(E\omega)^{2}}{m^{4}}
\eeq
where \,$(\omega_{1} - \omega_{2})$\, is taken to be positive. Taking also
\beq \label{Sp13}
d^{3}p_{1}\,\sim\,T^{3}\,,\,\,\,\,\,d^{3}q_{1}\,\sim\,m^{3}\,,\,\,\,\,\,q_{2}^{2}\,\sim\,m^{2}
\eeq
and doing an integration over $d^{3}p_{2}$ using \,$d^{3}q_{2}\,=\,q_{2}^{2}\,dq_{2}\,d\Omega_{q_{2}}$\, along with
\eq{Sp4}, \eq{Sp11}, \eq{Sp12} and \eq{Sp13}, one will get that the expression in \eq{Sp5} takes the form
\beq \label{Sp14}
\dot{\epsilon}^{el}\,\sim\,\frac{m_{0}^{5}}{\al}\Lb \frac{m}{m_{0}} \Rb^{5}(\al f_{q_{1}})(\al f_{q_{2}})
[(\omega_{1}-\omega_{2}) + T/f_{q_{2}} - T/f_{q_{1}}]\frac{1}{m}\,.
\eeq
From \eq{Sp3} and \eq{Sp4} we have
\bea \label{Sp15}
\frac{m}{m_{0}}\,\sim\,\left[2\,\al\,m_{0} \Lb \tau - \frac{\tau_{0}}
{\gamma}\ln\!\!\Lb \frac{1 + e^{\gamma(\tau/\tau_{0})}}{2} \Rb \Rb \right]^{1/4}\,.
\eea
As well as from \eq{Sp9} and \eq{Sp4} one obtains
\beq \label{Sp16}
\al f_{q}\,\sim\,\al T/\omega\,\sim\,\sqrt{\al}\,.
\eeq
Thus
\beq \label{Sp17}
\dot{\epsilon}^{el}\,\sim\,\frac{m_{0}^{5}}{\al}\frac{2}{1 + e^{\gamma(\tau/\tau_{0})}}\,\al^{9/4}\left[
m_{0}\Psi(\gamma,\tau)\right]^{5/4}[(\omega_{1}-\omega_{2}) + T/f_{q_{2}} - T/f_{q_{1}}]\,\frac{1}{m}
\eeq
where
\beq \label{Sp18}
\Psi(\gamma,\tau)\,\equiv\,2\,{\Large{\Lb \frac{1 + e^{\gamma(\tau/\tau_{0})}}{2}
\Rb^{4/5}}}\!\Lb \tau - \frac{\tau_{0}}{\gamma}\ln\!\!\Lb \frac{1 + e^{\gamma(\tau/\tau_{0})}}{2} \Rb \Rb\,.
\eeq
If $f_{q_{1}}$ and $f_{q_{2}}$ were take the form of \eq{Sp9} then we would obtain zero in the r.h.s. of \eq{Sp17}.
But the incoming flux of the soft gluons increases $f_{q}$ a little, so that the expression
\,$[(\omega_{1}-\omega_{2}) + T/f_{q_{2}} - T/f_{q_{1}}]$\, is greater than zero. However, from \eq{Sp17} it is obvious
that by the time \,$m_{0}\Psi(\gamma,\tau) \gg \al^{-9/5}$,\, $f_{q}$ will be very close to the thermal equilibrium distribution
making the \,$[\,\,]\frac{1}{m}$\, small and compensating the expression \,$\al^{9/4}[m_{0}\Psi(\gamma,\tau)]^{5/4}\,.$

Thus the energy is directly transferred from the incoming gluons of the source to the gluons having the momentum
$p\,\sim\,T$. Meanwhile, there is also an interaction with gluons having a momentum $k$ which is greater than the
soft scale $m$ and less than the hard scale $T$, but the flow from the source to the $k$-particles is suppressed by a factor
$k/T$ compared to the flow to the $T$-particles, so that by this way only a small fraction of the energy is transferred.

Consider the inelastic $3\,\leftrightarrow\,2$ scattering process (see \fig{inelastic}) which is shown to be
parametrically equally important as the elastic scattering. By this process the energy  is also transferred from the
scale $m$ to the scale $T$. The inelastic rate is
{\setlength\arraycolsep{2pt}
\bea \label{Sp19}
\dot{\epsilon}^{inel} & = & 2\,\frac{[2(N_{c}^{2}-1)]^{2}}{(2\pi)^{15}}\int_{R}\frac{d^{3}p_{1}}{2E_{1}}\frac{d^{3}p_{2}}{2E_{2}}
\frac{d^{3}p_{3}}{2E_{3}}\frac{d^{3}p_{4}}{2E_{4}}\frac{d^{3}k}{2E_{k}}\,(2\pi)^{4}\delta^{4}(p_{1} + p_{2} + k -
p_{3} - p_{4})\,|M_{12k\rightarrow34}|^{2}\,\times
\nonumber\\
& \times &
\,[f_{p_{1}}f_{p_{2}}f_{k}(1 + f_{p_{3}})(1 + f_{p_{4}}) - f_{p_{3}}f_{p_{4}}(1 + f_{p_{1}})(1 + f_{p_{2}})(1 + f_{k})]\,
\omega\,.
\eea}

\FIGURE{
\centerline{\epsfig{file=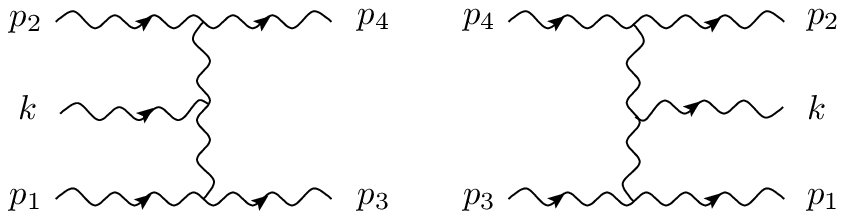,width=100mm,height=30mm}}
\caption{$3\,\leftrightarrow\,2$ inelastic scattering process. The figure is from \cite{MuShoWo06}.}
\label{inelastic} }

\hskip -0.7truecm
For a detailed calculation of this integral we refer to Ref.\,\cite{Wo04} as well. Here we put down the final result
\beq \label{sp20}
\dot{\epsilon}^{inel}\,\sim\, \al^{3}\frac{T^{6}}{m^{2}}\,[f_{p_{1}}f_{p_{2}}f_{k}(1 + f_{p_{3}})(1 + f_{p_{4}}) -
f_{p_{3}}f_{p_{4}}(1 + f_{p_{1}})(1 + f_{p_{2}})(1 + f_{k})]\,\omega\,.
\eeq
Using \eq{Sp4}, \eq{Sp9}, \eq{Sp10} and \eq{Sp15} one can rewrite \eq{sp20} as
\beq \label{Sp21}
\dot{\epsilon}^{inel} \sim \frac{m_{0}^{5}}{\al}\frac{2}{1 + e^{\gamma(\tau/\tau_{0})}}\,\al^{9/4}\left[
m_{0}\Psi(\gamma,\tau)\right]^{5/4}[\omega - T/f_{k}]\,\frac{1}{m}\,.
\eeq
It is clear that this equation is parametrically equivalent to \eq{Sp17}.

\subsection{The spectrum for\,\, $\al^{-7/5}\,<\,m_{0}\Psi(\gamma,\tau)\,<\,\al^{-9/5}$}
From \eq{Sp17} one clarifies that so long as \,$m_{0}\Psi(\gamma,\tau)\,<\,\al^{-9/5}$\,  it is impossible to transfer the
energy fast enough from the source in the region \,$m\,<\,\omega\,<\,\bar{m}$\, to the momentum region close to the scale
$T$ if we use a near equilibrium distribution in the soft region. In this scenario the gluons from the source ``pile up''
in the region \,$m\,<\,\omega\,<\,\bar{m}$\, whereby the occupation number becomes large enough speeding up the rate of
the energy transfer from the scale $m$ to the scale $T$ such that the transfer can compensate the energy rate incoming
from the source.

When $\omega\,>\,\bar{m}$ the distribution of the gluons is near to the equilibrium distribution of \eq{Sp8}, nevertheless,
when $m\,<\,\omega\,<\,\bar{m}$ the distribution will be noticeably changed (see \fig{distribut}).
Consider gluons having high momenta but located in the region $m\,<\,\omega\,<\,\bar{m}$. They will elastically scatter
with the gluons of momentum on the order of $T$ and will lose the energy to the harder gluons as given by \eq{Sp14}
which means that a gluon at the point 1 will move to the point 3 increasing $f_{k_{3}}$. As regards $f_{k_{1}}$, it is
determined by the difference between the incoming gluons causing $f_{k_{1}}$ to increase and the scattering on the hard
gluons causing $f_{k_{1}}$ to decrease. Therefore, with decrease of $k$ the occupation number $f_{k}$ will increase rapidly
until it becomes large enough such that the gluons of the momentum $k$ are absorbed by the hard ones at the same rate which
is determined by the emission of the external source and by gluons which acquire the momentum $k$ after elastic scattering
on the hard gluons. The estimates of these rates are given below.

\FIGURE{
\centerline{\epsfig{file=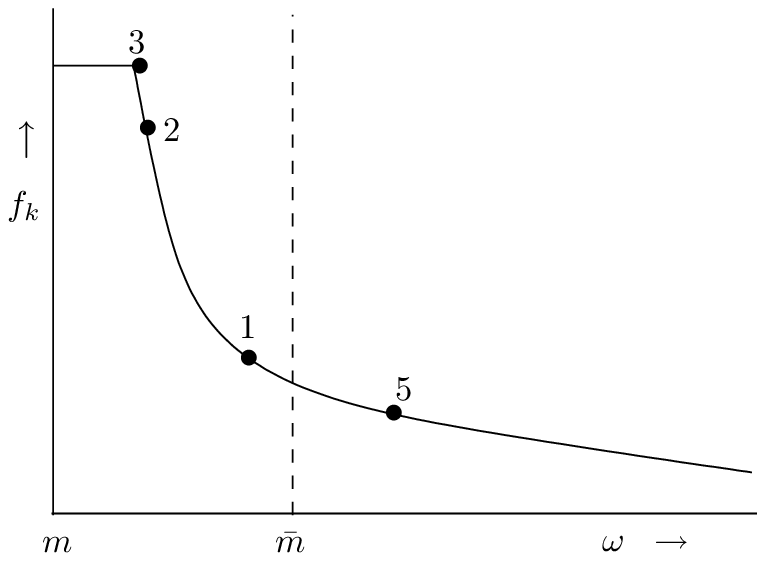,width=75mm,height=50mm}}
\caption{The gluon distribution as a function of the energy. The figure is from \cite{MuShoWo06}.}
\label{distribut}}

The case of the constant inflow of the energy, \,$\,\dot{\epsilon}\,=\,m_{0}^{5}/\al\,$,\, corresponds to an
increase in the occupation number, \,$\,\dot{f}_{k}\,\sim\,\frac{m}{\al}\Lb \frac{m_{0}}{m} \Rb^{5}$.
More precisely it comes from
\beq \label{Sp22}
\dot{\epsilon}\,\equiv\, \Lb 2(N_{c}^{2}-1)\int \frac{d^{3}k}{(2\pi)^{3}}\,\dot{f}_{k} \Rb \omega\,\,.
\eeq

\vskip -0.2truecm %
\hskip -0.7truecm %
To the time-dependent inflow of the energy, i.e., \,$\dot{\epsilon}\,=\,\frac{m_{0}^{5}}{\al}\,
\frac{2}{1 + e^{\gamma(\tau/\tau_{0})}}$\,,\, corresponds the increase in the occupation number, \,$\,\dot{f}_{k}\,\sim\,\frac{m}
{\al}\Lb\frac{m_{0}}{m} \Rb^{5}\!\!\frac{2}{1 + e^{\gamma(\tau/\tau_{0})}}$\,.

If we assume $k_{3}$ to be the momentum (see point 3 in \fig{distribut}) at which the loss by the inelastic absorption
exactly balances the rate of the incoming gluons directly from the source or coming from the source
via higher momentum regions then the incoming gluons rate will be
\beq \label{Sp23}
\dot{f}^{source}_{k_{3}}\,\sim\,\frac{1}{m^{4}}\frac{m_{0}^{5}}{\al}\frac{2}{1 + e^{\gamma(\tau/\tau_{0})}}\Lb
\frac{m}{k_{3}}\Rb^{3}
\eeq
where \,$\frac{1}{m^{4}}\frac{m_{0}^{5}}{\al}\frac{2}{1 + e^{\gamma(\tau/\tau_{0})}}$\, is the rate of the
total number of the incoming gluons arriving over the whole phase space \,$m\,<\,\omega\,<\,\bar{m}$. The term
\,$\Lb \frac{m}{k_{3}} \Rb^{3}$ means that these gluons terminate in the restricted region of the phase space on the order
of $k_{3}^{3}$. Analogously to \eq{sp20} the rate at which $k_{3}$-gluons are absorbed by the hard ones is (\fig{inelastic})
\beq \label{Sp24}
\dot{f}^{abs}_{k_{3}}\,\sim\,-\al^{3}\frac{T^{6}}{m^{5}}\left[f_{p_{1}}f_{p_{2}}f_{k_{3}}(1 + f_{p_{3}})
(1 + f_{p_{4}}) - f_{p_{3}}f_{p_{4}}(1 + f_{p_{1}})(1 + f_{p_{2}})(1 + f_{k_{3}})\right].
\eeq
Again using the procedure of Sec 3.1, \eq{Sp24} will lead to
\beq \label{Sp25}
\dot{f}^{abs}_{k_{3}}\,\sim\,-\frac{1}{m^{4}}\frac{m_{0}^{5}}{\al}\frac{2}{1 + e^{\gamma(\tau/\tau_{0})}}\,
\al^{9/4}\left[m_{0}\Psi(\gamma,\tau)\right]^{5/4}\sqrt{\al}f_{k_{3}}\,[\omega - T/f_{k_{3}}]\,\frac{1}{m}\,.
\eeq
Neglecting the term $1/f_{k_{3}}$ in this equation gives
\beq \label{Sp26}
\dot{f}^{abs}_{k_{3}}\,\sim\,-\frac{1}{m^{4}}\frac{m_{0}^{5}}{\al}\frac{2}{1 + e^{\gamma(\tau/\tau_{0})}}\,
\al^{9/4}\left[m_{0}\Psi(\gamma,\tau)\right]^{5/4}\sqrt{\al}f_{k_{3}}
\eeq
where it is conjectured that $k_{3}/m \ll 1$ taking $\omega_{3} \approx m$.

It should also be required that the $k_{3}$-gluons rate cascading to smaller momenta must not be large compared to those in
\eq{Sp23} and \eq{Sp26}. Considering the elastic scattering of the soft gluons, $k_{3}$ and \,$k_{4}$ ($k_{3}\,>\,k_{4}$),\,
with the hard gluons having momenta $p_{1}$, $p_{2}$ on the order of $T$, as shown in \fig{elastic} with the
replacement \,$q_{1}\,\rightarrow\,k_{3}$\, and \,$q_{2}\,\rightarrow\,k_{4}$, gives
{\setlength\arraycolsep{2pt}
\bea \label{Sp27}
\dot{f}_{k_{3}}^{3\rightarrow4} & = & -\frac{2(N_{c}^{2}-1)}{(2\pi)^{9}\,2\omega_{3}}\int\frac{d^{3}p_{1}}
{2E_{1}}\frac{d^{3}p_{2}}{2E_{2}}\frac{d^{3}k_{4}}{2\omega_{4}}\,(2\pi)^{4}\delta^{4}(p_{1} + k_{3} - p_{2} - k_{4})
\,|M|^{2}\,\times
\nonumber\\
& \times &
\,[f_{p_{1}}f_{k_{3}}(1 + f_{p_{2}})(1 + f_{k_{4}}) - f_{p_{2}}f_{k_{4}}(1 + f_{p_{1}})(1 + f_{k_{3}})]
\eea}
\hskip -0.1truecm
wherefrom
\beq \label{Sp28}
\dot{f}_{k_{3}}^{3\rightarrow4}\,\sim\,-\al^{2}\,\frac{T^{3}}{m^{4}}\,k_{4}^{3}\frac{1}{k_{3}}\left[f_{p_{1}}f_{k_{3}}
(1 + f_{p_{2}})(1 + f_{k_{4}}) - f_{p_{2}}f_{k_{4}}(1 + f_{p_{1}})(1 + f_{k_{3}})\right]
\eeq
which, in turn, yields
{\setlength\arraycolsep{2pt}
\bea \label{Sp29}
\dot{f}_{k_{3}}^{3\rightarrow4} & \sim & -\frac{1}{m^{4}}\frac{m_{0}^{5}}{\al}\frac{2}{1 + e^{\gamma(\tau/\tau_{0})}}\,
\al^{9/4}\left[m_{0}\Psi(\gamma,\tau)\right]^{5/4}\sqrt{\al}f_{k_{3}}\sqrt{\al}f_{k_{4}}\,\times
\nonumber\\
& \times &
\Lb \frac{k_{4}}{m} \Rb^{3}\Lb \frac{m}{k_{3}} \Rb [(\omega_{3} - \omega_{4}) + T/f_{k_{4}} - T/f_{k_{3}}]\,\frac{1}{m}.
\eea}
$\!\!$Making use of $\,\,\omega_{3}\,-\,\omega_{4} \sim \frac{k_{3}^{2}}{2m}\,\,$ as $\,\,k_{4} \leq k_{3} \ll m\,\,$
and neglecting the terms $1/f_{k_{3}}$ and $1/f_{k_{4}}$, \eq{Sp29} reduces to
\beq \label{Sp30}
\dot{f}_{k_{3}}^{3\rightarrow4}\,\sim\,-\frac{1}{m^{4}}\frac{m_{0}^{5}}{\al}\frac{2}{1 + e^{\gamma(\tau/\tau_{0})}}
\,\al^{9/4}\left[m_{0}\Psi(\gamma,\tau)\right]^{5/4}\sqrt{\al}f_{k_{3}}\sqrt{\al}f_{k_{4}}\Lb
\frac{k_{4}}{m} \Rb^{3}\Lb \frac{k_{3}}{m} \Rb\,.
\eeq
If we wish to find the value of the momentum $k_{3}$ at which the cascading to smaller momenta stops then it is
reasonable to suppose that \eq{Sp23}, \eq{Sp26} and \eq{Sp30} be of the same size when, in turn, the momenta $k_{3}$
and $k_{4}$ have the same sizes. So that using \eq{Sp15} and the conditions
\beq \label{Sp31}
|\,\dot{f}^{abs}_{k_{3}}|\,\sim\,|\,\dot{f}_{k_{3}}^{3\rightarrow4}|\,,
\eeq
\beq \label{Sp32}
|\,\dot{f}^{abs}_{k_{3}}|\,\sim\,|\,\dot{f}_{k_{3}}^{source}|
\eeq
one obtains
\beq \label{Sp33}
\frac{k_{3}}{m}\,\sim\,\Lb \al^{9/5}\left[m_{0}\Psi(\gamma,\tau)\right]\Rb^{5/4}\,,
\eeq
\beq \label{Sp34}
\sqrt{\al}f_{k_{3}}\,\sim\,\Lb \al^{9/5}\left[m_{0}\Psi(\gamma,\tau)\right]\Rb^{-5}\,.
\eeq

One must consider as well transitions due to the elastic scatterings when, first, $\omega$ goes from above $\bar{m}$ to
the point occupied by the $k_{3}$-gluons and, second, $\omega$ goes from below but near $\bar{m}$ to the point $k_{3}$.
Then we will have respectively
{\setlength\arraycolsep{2pt}
\bea \label{Sp35}
\dot{f}_{k_{5}}^{5\rightarrow3} & = & -\frac{2(N_{c}^{2}-1)}{(2\pi)^{9}\,2\omega_{5}}\int\frac{d^{3}p_{1}}{2E_{1}}
\frac{d^{3}p_{2}}{2E_{2}}\frac{d^{3}k_{3}}{2\omega_{3}}\,(2\pi)^{4}\delta^{4}(p_{1} + k_{5} - p_{2} - k_{3})\,|M|^{2}\,\times
\nonumber\\
& \times &
\,[f_{p_{1}}f_{k_{5}}(1 + f_{p_{2}})(1 + f_{k_{3}}) - f_{p_{2}}f_{k_{3}}(1 + f_{p_{1}})(1 + f_{k_{5}})]\,\,\,\Longrightarrow
\eea}

\vskip -1truecm
{\setlength\arraycolsep{2pt}
\bea \label{Sp36}
\Longrightarrow\,\,\,\dot{f}_{k_{5}}^{5\rightarrow3} & \sim & -\frac{1}{m^{4}}\frac{m_{0}^{5}}{\al}
\frac{2}{1 + e^{\gamma(\tau/\tau_{0})}}\,\al^{9/4}\left[m_{0}\Psi(\gamma,\tau)\right]^{5/4}\sqrt{\al}
f_{k_{3}}\sqrt{\al}f_{k_{5}}\,\times
\nonumber\\
& \times & \Lb \frac{k_{3}}{m} \Rb^{3} \Lb \frac{m}{k_{5}} \Rb
[(\omega_{5} - \omega_{3}) + T/f_{k_{3}} -
T/f_{k_{5}}]\,\frac{1}{m}
\eea}
and
{\setlength\arraycolsep{2pt}
\bea \label{Sp37}
\dot{f}_{k_{3}}^{1\rightarrow3} & = & \frac{2(N_{c}^{2}-1)}{(2\pi)^{9}\,2\omega_{3}}\int\frac{d^{3}p_{1}}{2E_{1}}
\frac{d^{3}p_{2}}{2E_{2}}\frac{d^{3}k_{1}}{2\omega_{1}}\,(2\pi)^{4}\delta^{4}(p_{1} + k_{1} - p_{2} - k_{3})\,|M|^{2}\,\times
\nonumber\\
& \times &
\,[f_{p_{1}}f_{k_{1}}(1 + f_{p_{2}})(1 + f_{k_{3}}) - f_{p_{2}}f_{k_{3}}(1 + f_{p_{1}})(1 + f_{k_{1}})]\,\,\,\Longrightarrow
\eea}

\vskip -1truecm
{\setlength\arraycolsep{2pt}
\bea \label{Sp38}
\Longrightarrow\,\,\,\dot{f}_{k_{3}}^{1\rightarrow3} & \sim & \frac{1}{m^{4}}\frac{m_{0}^{5}}{\al}\frac{2}{1 +
e^{\gamma(\tau/\tau_{0})}}\,\al^{9/4}\left[m_{0}\Psi(\gamma,\tau)\right]^{5/4}\sqrt{\al}f_{k_{1}}
\sqrt{\al}f_{k_{3}}\,\times
\nonumber\\
& \times &
\Lb \frac{k_{1}}{m} \Rb^{2} [(\omega_{1} - \omega_{3}) - T/f_{k_{1}}]\,\frac{1}{m}\,.
\eea}
For \,$1/f_{k_{3}}\,\ll\,1/f_{k_{5}}$,\, \,$m/k_{5}\,\sim\,1$\, and \,$\omega_{3}\,\approx\,m$\,  \eq{Sp36}
reduces to the following result:
{\setlength\arraycolsep{2pt}
\bea \label{Sp39}
\dot{f}_{k_{5}}^{5\rightarrow3} & \sim & -\frac{1}{m^{4}}\frac{m_{0}^{5}}{\al} \frac{2}{1 + e^{\gamma(\tau/\tau_{0})}}
\,\al^{9/4}\left[m_{0}\Psi(\gamma,\tau)\right]^{5/4}\sqrt{\al}f_{k_{3}}\sqrt{\al}f_{k_{5}}\,\times
\nonumber\\
& \times & \Lb \frac{k_{3}}{m} \Rb^{3}[(\omega_{5} - m) - T/f_{k_{5}}]\,\frac{1}{m}\,.
\eea}
$\dot{f}_{k_{5}}^{5\rightarrow3}$\, can be small in case of
\beq \label{Sp40}
f_{k_{5}}\,\simeq\,\frac{T}{\omega_{5} - m}
\eeq
taking into account that the term \,$\sqrt{\al}f_{k_{3}}\!\Lb \frac{k_{3}}{m} \Rb^{3}$\, is very large. But in this case
the occupation number is close to the thermal curve which means that \,$\sqrt{\al}f_{k_{5}}\,\sim\,1$\, (see \eq{Sp16})
and there can be no strong change in $f_{k}$ as $\omega$ passes the value $\bar{m}$ from above. As regards
\eq{Sp38}, for that case \,$(k_{1}/m)^{2}\,\sim\,1$\, which reduces the formula to
\beq \label{Sp41}
\dot{f}_{k_{3}}^{1\rightarrow3}\,\sim\,\frac{1}{m^{4}}\frac{m_{0}^{5}}{\al}\frac{2}{1 + e^{\gamma(\tau/\tau_{0})}}\,
\al^{9/4}\left[m_{0}\Psi(\gamma,\tau)\right]^{5/4}\sqrt{\al}f_{k_{1}}\sqrt{\al}f_{k_{3}}\,
[(\omega_{1} - m) - T/f_{k_{1}}]\,\frac{1}{m}\,.
\eeq
From comparison with \eq{Sp26} it is clear that $\sqrt{\al}f_{k_{1}}$ cannot be large since these two
equations are of the same order.

The results of this section and Sec 3.1 are similar to the corresponding results of Ref.\,\cite{MuShoWo06}. Basing on the
time-dependent law of the incoming energy rate we derived the modified Kolmogorov gluon spectra for the late and
intermediate time stages.

\subsection{The spectrum for\,\, $1\,\ll\,m_{0}\Psi(\gamma,\tau)\,<\,\al^{-7/5}$}
In \cite{MuShoWo06} it was shown that in the early time domain, \,$1\,\ll\,m_{0} \tau\,<\,\al^{-7/5}$,\, the system is far
from the thermal equilibrium in both high and low momentum regimes. Thereat $p_{0}(\tau)$ is the maximum scale to which the
gluons have evolved with \,$f_{p_{0}} \gg 1$,\, meanwhile, in the domain \,$m\,\ll\,\omega\,\ll\,p_{0}$\, the occupation
number is \,$f_{k}\,=\,\frac{c(\tau)}{\al}\frac{m}{\omega}$.\, The latter is a kind of an equilibrium distribution although
at the scale $p_{0}(\tau)$ it does not match onto the exact equilibrium distribution. From \eq{Sp34} one observes that in the
domain \,$m\,<\,\omega\,<\,\bar{m}$\, so long as \,$m_{0}\Psi(\gamma,\tau)$\, decreases from the point \,$\al^{-7/5}$,\,
\,$f_{k}$\, grows continuously, however, at \,$m_{0}\Psi(\gamma,\tau)\,\approx\,1$\, the occupation number is received much
more bigger than unity which is not consistent and significantly exceeds the case which we find below.

In this section we use another procedure of \cite{MuShoWo06} to obtain the gluon spectrum in the early time stage.
Suppose that at a time $\tau$ the gluon spectrum has reached the momentum $p_{0}(\tau)$ and let us remind that at
\,$\gamma\,=\,0$\, the energy conservation gives
\beq \label{Sp42}
\frac{m_{0}^{5}}{\al}\,\tau\,\sim\,f_{p_{0}}\,p_{0}^{4}
\eeq
and making use of
\beq \label{Sp43}
m^{2}\,\sim\,\al f_{p_{0}}\,p_{0}^{2}\,\,\,\,\,\,\,\,\,\mbox{gives}\,\,\,\,\,\,\,\,\Rightarrow\,\,\,\,\,\,\,\,
p_{0} m\,\sim\,m_{0}^{2}\sqrt{m_{0} \tau}.
\eeq
In case of \,$k/m\,\gg\,1$\, \,$f_{k}$\, is expected to have the form
\beq \label{Sp44}
f_{k}\,\simeq\,\frac{c(\tau)}{\al}\frac{m}{\omega}
\eeq
with \,$c(\tau),$\, \,$p_{0}$\, and \,$m$\, estimated to be as
\beq \label{Sp45}
c(\tau)\,\sim\,\frac{m}{p_{0}}\,\sim\,(m_{0} \tau)^{-5/14}\,,
\eeq
\beq \label{Sp46}
m\,\sim\,m_{0}\,(m_{0}\tau)^{1/14}\,,
\eeq
\beq \label{Sp47}
p_{0}\,\sim\,m_{0}\,(m_{0}\tau)^{3/7}\,.
\eeq

\hskip -0.7truecm
In our case the energy conservation gives
\beq \label{Sp48}
\frac{m_{0}^{5}}{\al}\,\bar{\tau}\,\sim\,f_{p_{0}}\,p_{0}^{4}
\eeq
where
\beq \label{Sp49}
\bar{\tau}\,\equiv\,2\Lb \tau - \frac{\tau_{0}}{\gamma}\ln\!\!\Lb
\frac{1 + e^{\gamma(\tau/\tau_{0})}}{2} \Rb \Rb
\eeq
(see \eq{Sp2});
then \eq{Sp43} and \eq{Sp44} will take the forms
\beq \label{Sp50}
p_{0} m\,\sim\,m_{0}^{2}\sqrt{m_{0}\bar{\tau}}\,,
\eeq
\beq \label{Sp51}
f_{k}\,\simeq\,\frac{c(\bar{\tau})}{\al}\frac{m}{\omega}\,.
\eeq
We conjecture for \,$c(\tau)$\, to be of the same form as \eq{Sp45}
\beq \label{Sp52}
c(\bar{\tau})\,\sim\,[m_{0}\Psi(\gamma,\tau)]^{-5/14}
\eeq
which is obtained \,\,(from a power dependence  $\,-\,\,c(\bar{\tau})\sim\al^{a}[m_{0}\Psi]^{b}$)\,\,
requiring \,\,$f_{k}(\tau)\,\sim\,1/\al$\,\, when \,$m_{0}\Psi(\gamma,\tau)\,\sim\,1$\, and requiring
\,$f_{k}(\tau)\,\sim\,1/\sqrt{\al}$\, when \,$m_{0}\Psi(\gamma,\tau)\,\sim\,\al^{-7/5}$,\, as given
by \eq{Sp3} and \,$f_{k}\,\simeq\,T/k$ \footnote{When \,$k/m\,\gg\,1$\, we will not distinguish between $\omega$ and $k$
in \eq{Sp9}.}.\, Using \eq{Sp51}, \eq{Sp52} and the expression
\beq \label{Sp53}
m^{2}\,\sim\,\al f_{p_{0}}\,p_{0}^{2}
\eeq
we arrive at
\beq \label{Sp54}
\frac{m}{p_{0}}\,\sim\,c(\bar{\tau})\,\sim\,[m_{0}\Psi(\gamma,\tau)]^{-5/14}
\eeq
while for \,$m$\, and \,$p_{0}$,\, \eq{Sp50} and \eq{Sp52} yield
\beq \label{Sp55}
m\,\sim\,m_{0}(m_{0}\bar{\tau})^{1/4}[m_{0}\Psi(\gamma,\tau)]^{-5/28}\,,
\eeq
\beq \label{Sp56}
p_{0}\,\sim\,m_{0}(m_{0}\bar{\tau})^{1/4}[m_{0}\Psi(\gamma,\tau)]^{5/28}\,.
\eeq
In the region \,$m\,<\,\omega\,<\,\bar{m}$\, the situation is almost like to the case considered in Sec. 3.2.
\beq \label{Sp57}
\dot{f}^{source}_{k_{3}}\,\sim\,\frac{1}{m^{4}}\frac{m_{0}^{5}}{\al}\frac{2}{1 + e^{\gamma(\tau/\tau_{0})}}\Lb
\frac{m}{k_{3}}\Rb^{3}\,,
\eeq
\beq \label{Sp58}
\dot{f}^{abs}_{k_{3}}\,\sim\,-\frac{1}{m^{4}}\frac{m_{0}^{5}}{\al}\frac{2}{1 + e^{\gamma(\tau/\tau_{0})}}\,
\al^{9/4}\left[m_{0}\Psi(\gamma,\tau)\right]^{5/4}\sqrt{\al}f_{k_{3}}\Lb \frac{1}{\sqrt{\al}}
\frac{m}{p_{0}} \Rb\,,
\eeq
\beq \label{Sp59}
\dot{f}_{k_{3}}^{3\rightarrow4}\,\sim\,-\frac{1}{m^{4}}\frac{m_{0}^{5}}{\al}\frac{2}{1 + e^{\gamma(\tau/\tau_{0})}}
\,\al^{9/4}\left[m_{0}\Psi(\gamma,\tau)\right]^{5/4}\sqrt{\al}f_{k_{3}}\sqrt{\al}f_{k_{4}}\Lb
\frac{k_{4}}{m} \Rb^{3}\Lb \frac{k_{3}}{m} \Rb
\eeq
where the explanation of the meaning of these expressions is exactly the same as in Sec 3.2. Only \eq{Sp26}
is now changed to \eq{Sp58} which is realized under the same assumption as for getting Eq.\,(64) of \cite{MuShoWo06}.
Requiring that \,$\dot{f}_{k_{3}}$\, terms in these three expressions be of the same size gives
\beq \label{Sp60}
\frac{k_{3}}{m}\,\sim\,\Lb \frac{m}{p_{0}} \Rb^{2}\Lb \frac{m}{m_{0}} \Rb^{5}\,\Lb \frac{1 + e^{\gamma(\tau/\tau_{0})}}{2}
\Rb\,=\,[m_{0}\Psi(\gamma,\tau)]^{-5/14}
\eeq
and
\beq \label{Sp61}
f_{k_{3}}\,\sim\,\frac{1}{\al}\Lb \frac{p_{0}}{m} \Rb^{7} \Lb \frac{m_{0}}{m} \Rb^{20}\Lb \frac{2}
{1 + e^{\gamma(\tau/\tau_{0})}} \Rb^{4}\,=\,\frac{1}{\al}\,[m_{0}\Psi(\gamma,\tau)]^{15/14}\,.
\eeq
which now replace \eq{Sp33} and \eq{Sp34}. In addition, it should be pointed out that \eq{Sp60} and \eq{Sp61} agree with
\eq{Sp33} and \eq{Sp34} at \,$m_{0}\Psi(\gamma,\tau)\,=\,\al^{-7/5}$.

However, additionally, we wish to look for solutions with another assumption for \,$c(\bar{\tau})$,\, i.e., when
\beq \label{Sp62}
c(\bar{\tau})\,\equiv\,c\,'\!(\bar{\tau})\,\sim\,(m_{0} \bar{\tau})^{-5/14}
\eeq
which like the above case is obtained \,(from a power dependence  $\,-\,\,c(\bar{\tau})\sim\al^{a}(m_{0}\bar{\tau})^{b}$)\,\,
requiring \,$f_{k}(\tau)\,\sim\,1/\al$\, when \,$m_{0}\bar{\tau}\,\sim\,1$\, and requiring
\,$f_{k}(\tau)\,\sim\,1/\sqrt{\al}$\, when \,$m_{0}\bar{\tau}\,\sim\,\al^{-7/5}$,\, as given by \eq{Sp3} and
\,$f_{k}\,\simeq\,T/k$.\, Using \eq{Sp50} $-$ \eq{Sp53} one gets
\beq \label{Sp63}
\frac{m}{p_{0}}\,\sim\,c\,'\!(\bar{\tau})\,\sim\,(m_{0} \bar{\tau})^{-5/14}\,,
\eeq
\beq \label{Sp64}
m\,\sim\,m_{0}\,(m_{0}\bar{\tau})^{1/14}\,,
\eeq
\beq \label{Sp65}
p_{0}\,\sim\,m_{0}\,(m_{0}\bar{\tau})^{3/7}\,.
\eeq
Consequently, \eq{Sp57} $-$ \eq{Sp59} give
\beq \label{Sp66}
\frac{k_{3}}{m}\,\sim\,\Lb \Lb \frac{1 + e^{\gamma(\tau/\tau_{0})}}{2} \Rb^{-14/5} (m_{0}\bar{\tau}) \Rb^{-5/14}\,,
\eeq
\beq \label{Sp67}
f_{k_{3}}\,\sim\,\frac{1}{\al}\Lb \Lb \frac{1 + e^{\gamma(\tau/\tau_{0})}}{2} \Rb^{-56/15} (m_{0}\bar{\tau})
\Rb^{15/14}\,.
\eeq
But one must be aware of the fact that when $m_{0}$ is small and $\gamma$ has higher values, in this case
\,$m_{0} \bar{\tau}\,<\,1$,\, so that our assumption will bring to incorrect results. Moreover, the functions in the
parentheses in \eq{Sp66} and \eq{Sp67} reach to \,$m_{0}\bar{\tau}\,=\,\al^{-7/5}$\, only when \,$\gamma\,=\,0$.\,
Therefore, in such a situation we use somewhat naive procedure by which these functions agree with \eq{Sp33} and
\eq{Sp34} at \,$m_{0}\bar{\tau}\,=\,\al^{-7/5}$.\, Namely, one can do that performing the following interpolation:
\beq \label{Sp68}
\frac{k_{3}}{m}\,\sim\,\Lb \Lb \frac{1 + e^{\gamma(\tau/\tau_{0})}}{2} \Rb^{(-14/5) + (\xi\cdot\tau/5)} (m_{0}\bar{\tau})
\Rb^{-5/14}\,,
\eeq
\beq \label{Sp69}
f_{k_{3}}\,\sim\,\frac{1}{\al}\Lb \Lb \frac{1 + e^{\gamma(\tau/\tau_{0})}}{2} \Rb^{-(56/15) + (\xi'\cdot\tau/5)}
(m_{0}\bar{\tau})  \Rb^{15/14}
\eeq
where $\xi$ is chosen such that \,\,$\xi\cdot\tau_{\al^{7/5}}\,\equiv\,\xi(\tau)\cdot\tau_{\al^{7/5}}\,=\,18$\,\,
and \,\,$\xi'\cdot\tau_{\al^{7/5}}\,\equiv\,\xi'(\tau)\cdot\tau_{\al^{7/5}}\,\cong\,22.667$\,\,
\footnote{$\tau_{\al^{7/5}}$\, denotes the evolution time of the gluon system at \,$m_{0}\bar{\tau}\,=\,\al^{-7/5}$.}.

Simultaneously, we wish to notice that in the ``bottom-up'' approach when  $Q_{s}\tau$ is less than 1, it is more appropriate
to describe the gluon field as a nonlinear gluon field rather than a collection of the hard particles due to strong interactions
between the gluons. But when \,$Q_{s}\tau$\, becomes larger than 1, the gluons can be described as particles on mass shell.
In QCD wave turbulence we also consider the gluons on the mass shell. In \cite{MuShoWo06} this occurs when
\,$m_{0}\tau\,>\,1$.\, So that we accept that the gluons are on the mass shell in case of
\beq \label{Sp70}
Q_{s}\tau\,\sim\,m_{0}\tau\,>\,1\,\,\,\,\,\,\,\,\,\,\,\mbox{which has a lower boundary}\,\,\,\,\,\,\,\,\,\,
Q_{s}\tau_{tr}\,\sim\,m_{0}\tau_{tr}\,\sim\,1
\eeq
wherefrom \,$m_{0}\,\sim\,Q_{s}$.\, $\tau_{tr}$ is the transition time point from the non-linear description of the gluonic
field to the linear one. From this point of view, for RHIC energies \,$m_{0}\,\sim\,1 - 1.4\,GeV\,\sim\,5 - 7\,fm^{-1}$.
In our discussion when the flow weakening parameter $\gamma$ is non-zero, \eq{Sp70} switches over to
\beq \label{Sp71}
Q_{s}\tau\,\sim\,m_{0}\Psi(\gamma,\tau)\,>\,1\,\,\,\,\,\,\,\,\,\,\,\mbox{which has a lower boundary}\,\,\,\,\,\,\,\,\,\,
 Q_{s}\tau_{tr}\,\sim\,m_{0}\Psi(\gamma,\tau_{tr})\,\sim\,1\,.
\eeq

\FIGURE{
\centerline{\epsfig{file=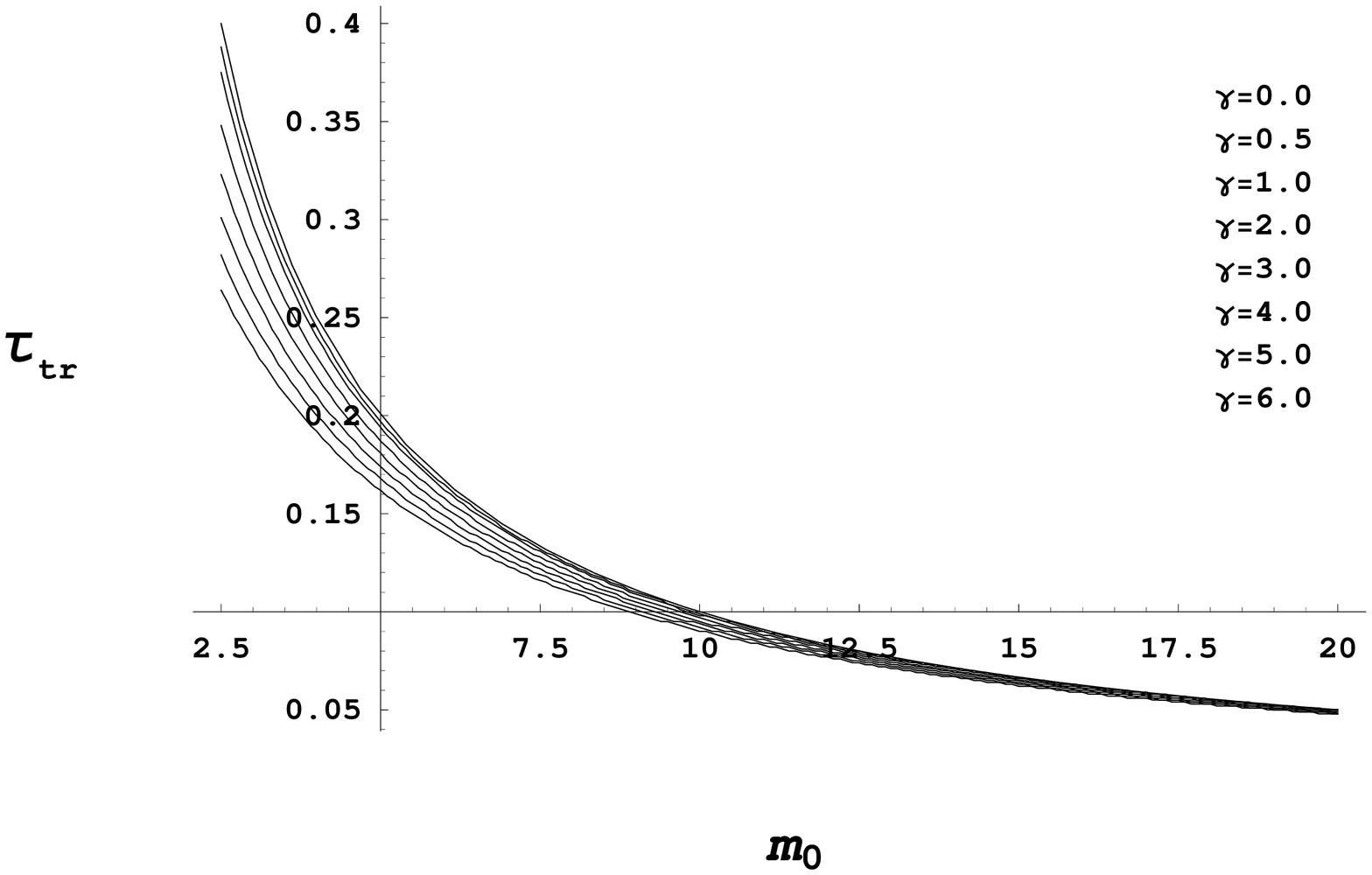,width=110mm,height=65mm}}
\caption{The transition time \,$\tau_{tr}$\, versus \,$m_{0}$\, from \,$m_{0}\Psi(\gamma,\tau_{tr})\,=\,1$ (see
\protect\eq{Sp61} and \protect\eq{Sp18}).\, The upper curve corresponds to \,$\gamma\,=\,0$\, and the other ones correspond to
the other \,$\gamma$'s\, from top to bottom respectively.}
\label{trans1} }

In \fig{trans1} one can see that for various $\gamma$'s the values of the transition time $\tau_{tr}$, at a fixed $m_{0}$, do
not differ from each other significantly, especially, at higher $m_{0}$'s. In \fig{trans_RHIC} and \fig{trans_LHC} we plot
the curves of the transition time versus \,$m_{0}$\, at \,$\al\,\approx\,0.43$\, and \,$\al\,\approx\,0.29$\, \footnote{
These values of $\al$ with the corresponding $Q_{s}$ are estimated in the next section.} fixed, respectively, for RHIC and
LHC energies. In \fig{trans_RHIC_LHC} we plot the curves of the average transition time versus \,$m_{0}$\, at
\,$\al\,\approx\,0.43$\, and \,$\al\,\approx\,0.29$\, by averaging the sum of two transition times, correspondingly,
obtained from the conditions \,$m_{0}\Psi(\gamma,\tau_{tr})\,=\,1$\, and $function\,=\,1$ \footnote{See the function in the
parenthesis of \protect\eq{Sp69}.}. In \fig{trans_RHIC} and \fig{trans_LHC} one can see that \,$\tau_{tr}$\, acquires
relatively bigger values at non-zero \,$\gamma$'s\, when we come from the assumption
\,$c\,'\!(\bar{\tau})\,\sim\,(m_{0} \bar{\tau})^{-5/14}$.\, But for increasing $\al$ the values of \,$\tau_{tr}$\, decrease.
Meanwhile, in \fig{trans1} one can see that \,$\tau_{tr}$\, acquires relatively smaller values with increase of
\,$\gamma$\, when we come from the assumption \,$c(\bar{\tau})\,\sim\,[m_{0}\Psi(\gamma,\tau)]^{-5/14}$.\, And if we
consider these two assumptions as limits of the domain where the accurate solution to the early time gluon spectrum exists
then we admit that the correct values of the transition time are located in the range between the upper and lower curves in
each of the panels of \fig{trans_RHIC_LHC}. Nonetheless, we prone to suppose that the accurate solution is close to
\fig{trans1} which depicts the solution in \eq{Sp60} and \eq{Sp61} wherefrom \,$\tau_{tr}$\, is equally defined, meanwhile,
\,$\tau_{tr}$\, received from \eq{Sp69} is \,$1\,\div\,1.35$\, times greater than received from \eq{Sp68} which shows that
the admission in \eq{Sp62} is simplified.

\FIGURE[ht]{
\centerline{\epsfig{file=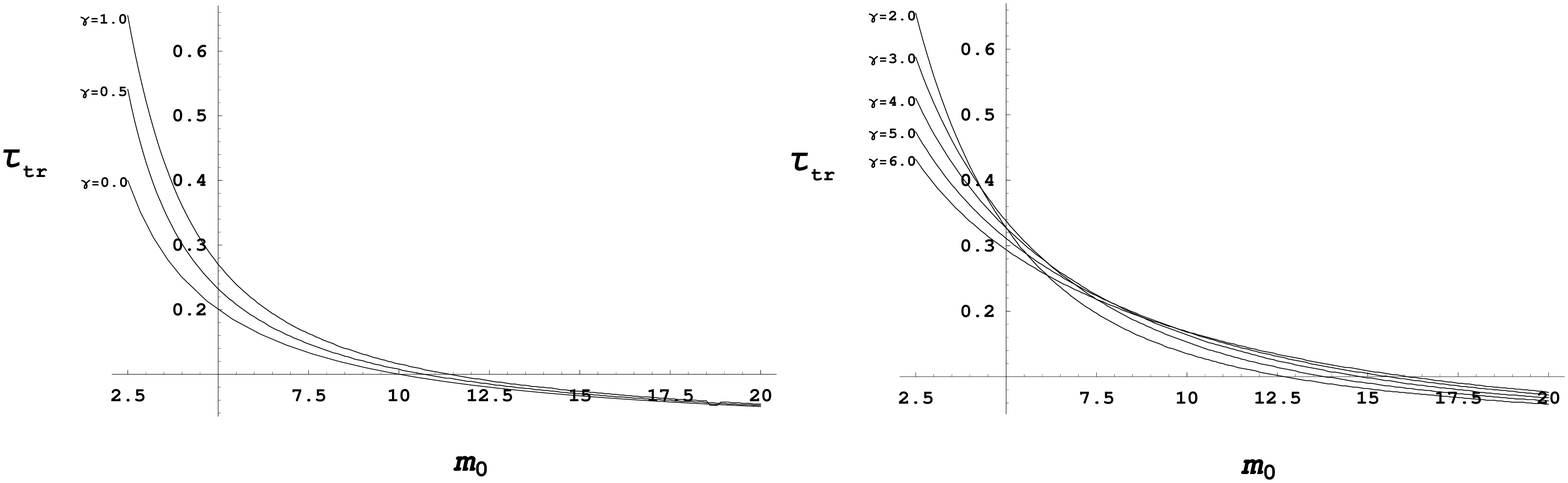,width=185mm,height=58mm}}
\caption{The transition time \,$\tau_{tr}$\, versus \,$m_{0}$\, for smaller \,$\gamma$'s\, (left panel) and higher
\,$\gamma$'s\, (right panel) at \,$\al\,\approx\,0.43$\, from \protect\eq{Sp69}.}
\label{trans_RHIC} }

\FIGURE[ht]{
\centerline{\epsfig{file=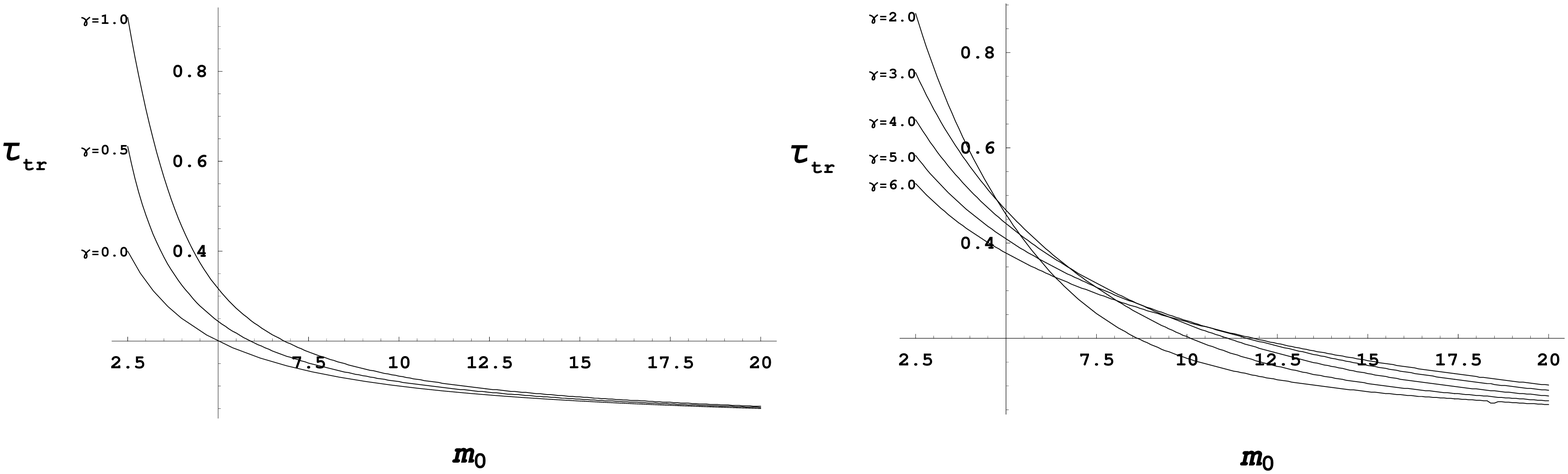,width=185mm,height=58mm}}
\caption{The transition time \,$\tau_{tr}$\, versus \,$m_{0}$\, for smaller \,$\gamma$'s\, (left panel) and higher
\,$\gamma$'s\, (right panel) at \,$\al\,\approx\,0.29$\, from \protect\eq{Sp69}.}
\label{trans_LHC} }

\FIGURE[ht]{
\centerline{\epsfig{file=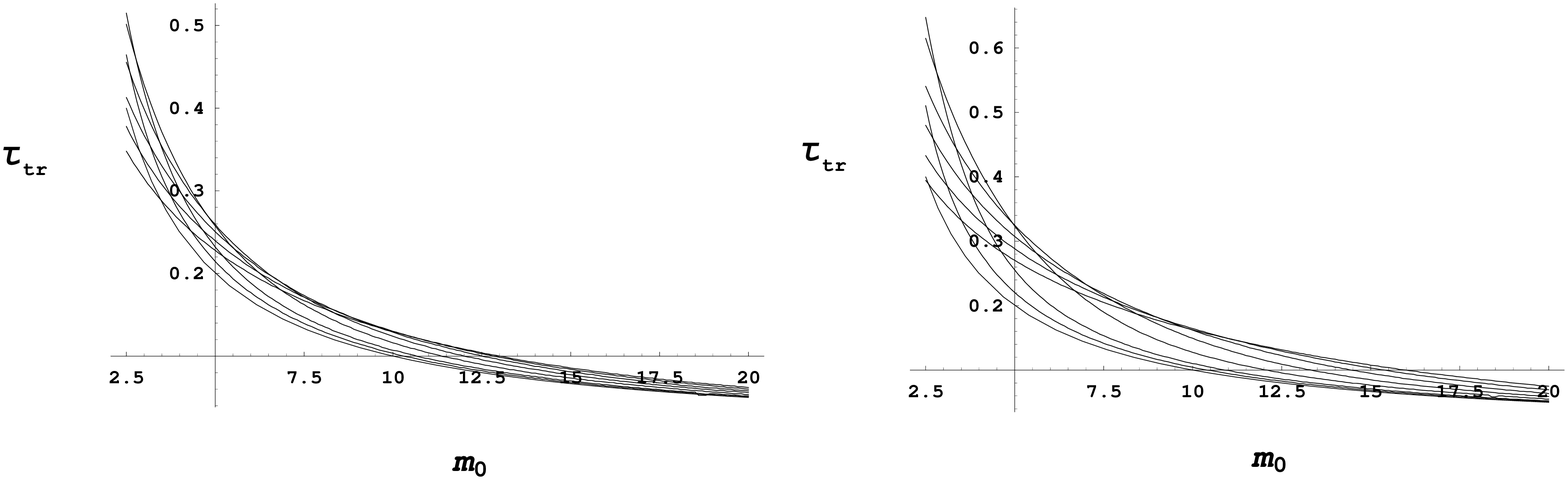,width=185mm,height=58mm}}
\caption{The transition time \,$\tau_{tr}$\, versus \,$m_{0}$\, at \,$\al\,\approx\,0.43$\, (left panel) and
\,$\al\,\approx\,0.29$\, (right panel) by averaging the sum of two transition times, correspondingly, obtained from
the conditions \,$m_{0}\Psi(\gamma,\tau_{tr})\,=\,1$\, and $function\,=\,1$ (in the parenthesis of \protect\eq{Sp69}).}
\label{trans_RHIC_LHC} }

In \cite{MuShoWo06} the mean free path, \,$\lambda_{p_{0}}$,\, of the gluons with momentum \,$p_{0}$\, has the
following form:
\beq \label{Sp72}
\frac{\lambda_{p_{0}}}{\tau}\,\sim\,\frac{1}{\tau p_{0}^{3} f_{p_{0}}^{2}\al^{2}/p_{0}^{2}}\,=\,\frac{p_{0}^{3}}{m^{4}
\tau}\,\sim\,1
\eeq
which demonstrates why \eq{Sp45} must hold. If \,$c(\tau)$\, were to be parametrically larger than that given by \eq{Sp45}
then \,$\lambda_{p_{0}}/\tau$\, would be less than one which, in turn, would bring to instability of the distribution in
\eq{Sp44}. If \,$c(\tau)$\, were to be much smaller than given by \eq{Sp45} then the maximum value of $f_{k}$, for
very small values of $k$, would vastly exceed that given in \eq{Sp61} at \,$\gamma\,=\,0$.
The r.h.s. of \eq{Sp72} is derived using \eq{Sp46} and \eq{Sp47} which corresponds to \,$\gamma\,=\,0$.\, If we use
\eq{Sp64} and \eq{Sp65}, having \,$c(\bar{\tau})$\, in \eq{Sp63}, the result again is the same as in \eq{Sp72}, i.e.,
\,$\lambda_{p_{0}}/\tau\,\sim\,1$. From the other side when one uses \eq{Sp55} and \eq{Sp56}, having \,$c(\bar{\tau})$\, in
\eq{Sp54}, the result will approach to unity at \,$\tau\,\rightarrow\,\tau_{tr}$\, (where \,$\tau_{tr}$\, is defined
from \,$m_{0}\Psi(\gamma,\tau_{tr})\,\sim\,1$):
\beq \label{Sp73}
\frac{\lambda_{p_{0}}}{\tau}\,\sim\,\frac{1}{\tau p_{0}^{3} f_{p_{0}}^{2}\al^{2}/p_{0}^{2}}\,=\,\frac{p_{0}^{3}}{m^{4}
\tau}\,=\,\frac{(m_{0}\Psi)^{5/4}}{(m_{0}\tau)\,(m_{0}\bar{\tau})^{1/4}}\,\Rightarrow
\eeq
\beq \label{Sp74}
\Rightarrow\,\frac{\lambda_{p_{0}}}{\tau}\,\sim\, \left\{ \begin{array}{ll}
1\div\,1.4\,\,\,\,(\mbox{for}\,\,\gamma\,=\,0\,\div\,6)\,\,\,\,\mbox{at}\,\,\,\,m_{0}\,\sim\,1\,GeV\,\sim\,5\,fm^{-1}\,
(\mbox{RHIC}),\\
1\div\,1.2\,\,\,\,(\mbox{for}\,\,\gamma\,=\,0\,\div\,6)\,\,\,\,\mbox{at}\,\,\,\,m_{0}\,\sim\,2.3\,GeV\,\sim\,11.6\,fm^{-1}\,
(\mbox{LHC}).
\end{array} \right.
\eeq
In this regard in \eq{Sp51} one can probably consider the distribution $f_{k}$ to be if not genuinely stable but rather
quasi-stable. Thus it is natural to suppose that the consistency and stability of \eq{Sp44} \footnote{See discussion in
Ref.\,\cite{MuShoWo06}.} is somewhat extended to our case for \eq{Sp51}.

\section{Matching of the equilibration times of QCD wave turbulence and ``bottom-up'' thermalization}
\label{sec:Matching}

From previous sections we know that
\begin{itemize}
\item[1)] In the ``bottom-up'' scenario the system of the gluons reaches the thermal equilibrium at
\,$Q_{s}\tau\,>\,\al^{-5/2}$\, when \,$Q_{s}\tau\,>\,\al^{-13/5}$;
\item[2)] In QCD wave turbulence scenario the system of the gluons reaches the thermal equilibrium at
\,$m_{0}\Psi(\gamma,\tau)\,\gg\,\al^{-9/5}$.
\end{itemize}
In this section we exhibit at what conditions the equilibration time of QCD wave turbulence can be matched onto $\tau_{eq}$
of the ``bottom-up'' thermalization. It will be clear that the previously derived Kolmogorov gluon spectra are such that
the resulting equilibration time can be matched onto $\tau_{eq}$ of various evolutional approaches as well. Thus we will have
the ``running'' with time gluonic spectra \footnote{We will see that the word ``running'' is due to the energy flow weakening
parameter $\gamma$.}.

\subsection{The picture at RHIC}
The solution to the Yang-Mills equation with a strong color source at rapidity \,$y = 0$\, and momenta \,$k_{T} \gg
\Lambda_{QCD}$\, $(\Lambda_{QCD} = 0.2\,GeV)$\, gives the nucleus gluon distribution \cite{JaKo,Ko96} of the form
\beq \label{Rh1}
\phi_{A}(k_{T})\,=\,\frac{N_{c}^{2} - 1}{4\pi^{4}\al N_{c}}\int \frac{d^{2}r_{T}^{}}{r_{T}^{2}}\,e^{-ik_{T}\cdot
r_{T}^{}} \Lb 1 - e^{-(r_{T}^{2}/4)\,Q_{s}^{2}(r_{T}^{2})} \Rb
\eeq
where $r_{T}^{}$-dependent saturation momentum is taken as follows \cite{Mu99,BaKoWi,KLN}:
\beq \label{Rh2}
Q_{s}^{2}(r_{T}^{2},b)\,=\,\frac{4\pi^{2}\al N_{c}}{N_{c}^{2} - 1}\,xG(x,1/r_{T}^{2})\frac{\rho_{part}(b)}{2}
\eeq
with \,$xG(x,1/r_{T}^{2})$\, being the gluon structure function of the nucleon and \,$\rho_{part}(b)$\, being the density of
participating nucleons in the transverse plane as a function of the impact parameter $b$ of $AA$ collisions.
The nucleon gluon distribution can be taken to be of the perturbative form \cite{BaKoWi,Mul1} as
\beq \label{Rh3}
xG(x,1/r_{T}^{2})\,=\,F\,\frac{\alpha_{s}(N_{c}^{2} - 1)}{2\pi}\,\ln\!\!{\Lb \frac{1}{r_{T}^{2}\Lambda_{QCD}^{2}} + cut \Rb}.
\eeq
Here the numerical multiplicative factor $F$ reflects the fact that at low-$x$ the gluons, along with initiation from the
valence quarks, additionally, originate from energetic gluons and sea quarks; \,$cut = \frac{1}{r_{T,cut}^{2}\Lambda_{QCD}^{2}}$\,
is the infrared cutoff, a small regulator providing the saturation momentum to be remained positive for
\,$r_{T}^{2}\,\gg\,r_{T,cut}^{2}$.\, We take \,$r_{T,cut}^{}$\, to be equal to \,$3\,GeV^{-1}$\, \cite{BaKoWi}, such that
for momenta \,$k_{T}\,\geq\,O(1\,GeV)$\, the sensitivity on $cut$ is negligible. The nuclear saturation scale
\,$Q_{s}^{2}(b)$\, is self-consistently obtained from the solution of \eq{Rh2} when evaluated at
\,$r_{T}^{2}\,=\,1/Q_{s}^{2}(b)$.\, In \cite{BaKoWi} \,$Q_{s}^{2}\,=\,2\,GeV^{2}$\, at \,$b = 0$\, and \,$\al = 0.5$\,
which gives \,$F = 1.8$\, in \eq{Rh3}.

In our paper we take \,$Q_{s}^{2}(b = 0)\,=\,1\,GeV^{2}$\, inspired by explanation of RHIC hadron multiplicities
in the ``bottom-up'' thermalization \cite{BMSS}.
For fixed $\al$ we employ a value calculated from the known expression of the coupling constant to one-loop order:
\beq \label{Rh4}
\al(Q_{s}^{2})\,\simeq\,\frac{4\pi}{\beta_{0}\ln(Q_{s}^{2}/\Lambda_{QCD}^{2})}
\eeq
with \,$\beta_{0}\,=\,(11 - 2n_{f}/3)$\, using \,$n_{f}\,=\,3$\, at \,$N_{c}\,=\,3$.\,
All these correspond to \,$F \simeq 1.4$\, in \eq{Rh3}.

It should be stressed that \,$Q_{s}^{2}\,=\,1\,GeV^{2}$\, at RHIC \,$W = 130\,GeV$\, is defined as a first approximation
from the expression
\beq \label{Rh5}
Q_{s}^{2}(b)\,\equiv\,\bar{Q}_{s}^{2}(b)\,=\,K\,\frac{4\pi^{2}N_{c}}{N_{c}^{2} - 1}\,\al (\bar{Q}_{s}^{2})\,
xG(x,\bar{Q}_{s}^{2})\,\frac{\rho_{part}(b)}{2}
\eeq
which stems from
\beq \label{Rh6}
Q_{s}^{2}(r_{T}^{2},b)\,=\,\frac{4\pi^{2}N_{c}}{N_{c}^{2} - 1}\,\al (Q_{s}^{2})\,xG(x,Q_{s}^{2})\,
\frac{\rho_{part}(r_{T}^{2},b)}{2}
\eeq
where $K$ is a multiplicative factor showing up as a consequence of the specific uncertainty in the exact determination of
the saturation momentum outside of the McLerran-Venugopalan model \cite{McVe}. This factor has influence on the average
transverse momentum per produced gluon but not on the total gluon number. As regards \,$\bar{Q}_{s}^{2}(b)$,\, it appears
as an effective average over the variable $r_{T}^{}$ in \eq{Rh6}. In \cite{BMSS} \,$\bar{Q}_{s}^{2}(b\,=\,0)$\, is
estimated to be \,$1\,GeV^{2}$\, for \,$K\,=\,1.6$\, at \,$\rho_{part}\,=\,3.06\,fm^{-2}$.

Since in the nucleon the color fields are weak, one can rely on the linear approximation. Then the structure function
\,$xG(x,Q_{s}^{2})$\, is governed by the DGLAP-like evolution \cite{KLN,DGLAP}:
\beq \label{Rh7}
xG(x,Q_{s}^{2})\,=\,0.42\,\ln\!\!\Lb \frac{Q_{s}^{2}}{\Lambda_{QCD}^{2}} \Rb\,.
\eeq
The constant factor is defined using the MRST parton distributions to next-to-next-leading-order (NNLO) in $\al$
\cite{MRST}. With \eq{Rh7} we get \,$K\,\simeq\,1.9$\, in \eq{Rh5}.

At this point it is convenient to change the exponential form of \eq{Rh1} making use of \eq{Rh3} and \eq{Rh5}:
\beq \label{Rh8}
\phi_{A}(k_{T})\,=\,\frac{N_{c}^{2} - 1}{4\pi^{4}\al N_{c}}\int \frac{d^{2}r_{T}^{}}{r_{T}^{2}}\,e^{-ik_{T}\cdot
r_{T}^{}}\left[ 1 - \exp\!\!\Lb -\frac{\,1}{\,4}\,r_{T}^{2}\,q^{2}\ln\!\!{\Lb \frac{1}{r_{T}^{2}\Lambda_{QCD}^{2}} + cut \Rb}
\Rb \right]
\eeq
where
\beq \label{Rh9}
q^{2}\,\equiv\,\Lb \frac{F}{0.42\,K} \Rb \frac{\al (N_{c}^{2} - 1)}{2\pi}\,\frac{Q_{s}^{2}}{\ln\!{(Q_{s}^{2}/
\Lambda_{QCD}^{2})}}\,.
\eeq
The upper cutoff of the integral is \,$r_{T,up}^{} = \frac{1}{\sqrt{1 - cut}\,\,\Lambda_{QCD}}$\, but as long as
\,$k_{T}\,\gg\,\Lambda_{QCD}$\, the value of the integral is very little sensitive to \,$r_{T,up}^{}$\,.\, In favor of the
selected saturation scale \,$Q_{s}^{2}(b = 0)\,=\,1\,GeV^{2}$\, used in \cite{BMSS}, from \eq{Rh3}, \eq{Rh5} and \eq{Rh7}
one can obtain
\beq \label{Rh10}
q^{2}\ln\!\!{\Lb \frac{1}{r_{T}^{2}\Lambda_{QCD}^{2}} + cut \Rb}\,=\,\frac{Q_{s}^{2}}{xG(x,Q_{s}^{2})}\,\cdot\,
\overline{xG(x,1/r_{T}^{2})}|_{r_{T}^{2}=1/Q_{s}^{2}}\,\simeq\,1\,GeV^{2}
\eeq
where
\beq \label{Rh11}
\overline{xG(x,1/r_{T}^{2})}|_{r_{T}^{2}=1/Q_{s}^{2}}\,\equiv\,(1/K)\cdot xG(x,1/r_{T}^{2})|_{r_{T}^{2}=1/Q_{s}^{2}}\,\simeq\,
xG(x,Q_{s}^{2})\,\,\,\,\,\,\,\mbox{at}\,\,\,\,\,Q_{s}^{2}\,=\,1\,GeV^{2}.
\eeq
So that by our opinion the selected parameters \,$F \simeq 1.4$\, in \eq{Rh3} and \,$K \simeq 1.9$\, in \eq{Rh5} give
self-consistently \,$Q_{s}^{2}(b = 0)\,\simeq\,1\,GeV^{2}$\, at RHIC, between the nucleon gluon perturbative distribution
and the gluon structure function governed by the DGLAP-like evolution.
Integrating the density of the participating nucleons with respect to $r_{T}^{}$ gives
\beq \label{Rh12}
\int d^{2}r_{T}^{}\,\rho_{part}(r_{T}^{2}, b)\,=\,N_{part}(b)
\eeq
then taking the rapidity distribution of freed gluons \,$dN/dy$\, at $y = 0$ and at given $b$, one can write
{\setlength\arraycolsep{2pt}
\bea \label{Rh13}
\frac{dN}{dy}(b) & = & \varepsilon\,\frac{N_{c}^{2}-1}{4\pi^{2} N_{c}}\int d^{2}r_{T}^{}\,\frac{1}{\al}\,Q_{s}^{2}
(r_{T}^{2},b)
\nonumber\\
\frac{dN}{dy}(b) & \simeq & \varepsilon\,\,x G(x, \bar{Q_{s}^{2}})\,\frac{N_{part}(b)}{2}
\eea}
$\!\!\!\!\!$ where the gluon liberation coefficient $\varepsilon$ accounts for the transformation of the gluons from the
initial into final state.

For the gluon structure function we take  \,$x G(x, Q^{2})\,\simeq\,1.65$\, at \,$Q^{2}\,=\,2\,GeV^{2}$\,  \cite{MRST}
(at \,$x\,\simeq\,0.02$\, and \,$W\,=\,130\,GeV$\, with \,$x\,=\,2 Q/W$). Then using \eq{Rh7} and the relation
\,\,$dN/dy\,\simeq\,(3/2)\,dN_{ch}/dy$\,\, \cite{KLN,Xi} one can rewrite \eq{Rh13} as follows:
\beq \label{Rh14}
\left<\frac{2}{N_{part}}\frac{dN_{ch}}{dy} \right>\,\simeq\,0.28\,\varepsilon\,\ln\!\!\Lb \frac{\bar{Q_{s}^{2}}}
{\Lambda_{QCD}^{2}} \Rb
\eeq
where the averaging in the l.h.s. is over events having different number of participants.

As noted in Sec 2.1, in the original ``bottom-up'' thermalization the hard gluons lose energy by radiating soft gluons in the
hard branching process. First, the hard gluon emits a gluon with a softer momentum, \,$k_{br}$,\, which splits into two gluons
with comparable momenta during a time $\tau$. The branching momentum was found to be \cite{BaMu}
\beq \label{Rh15}
k_{br}\,\sim\,\al^{4}T^{3}\tau^{2}\,.
\eeq
Then the products of this branching rapidly cascade further giving all their energy to the thermal bath formed by the soft
gluons. At the time \,$\tau\,\sim\,\al^{-5/2}Q_{s}^{-1}$\, the number of the soft gluons dominate that of the primary
hard gluons. Parametrically at \,$\tau\,\sim\,\al^{-13/5}Q_{s}^{-1}$\, the soft gluons system achieves the full
thermalization with the temperature of the order of \,$T\,\sim\,\al^{2/5}Q_{s}$.\,

In the {\em m}``bottom-up'' thermalization for gluons produced in the interval \,$Q_{s}^{-1} < \tau_{0} < \tau$,\, i.e.,
after the onset but before $\tau$ the scaling solutions are \cite{MuShoWo05}
$$
N_{s}(\tau,\tau_{0})\,\sim\,\frac{Q_{s}^3}{\al(Q_{s}\tau)(Q_{s}\tau_{0})^{1/3-\delta}}\,,\,\,\,\,\,\,\,\,\,\,\,\,\,\,\,
k_{s}(\tau_{0})\,\,\sim\,\,\frac{Q_{s}}{(Q_{s}\tau_{0})^{1/3 - 2\delta/5}}\,,
$$
\beq \label{Rh16}
\,\,\,\,\,\,\,\,\,\,\,\,\,\,\,\,\,\,\,\,
f_{s}(\tau,\tau_{0})\,\sim\,\frac{(Q_{s}\tau_{0})^{1/3 + \delta/5}}{\al(Q_{s}\tau)^{2/3 + 2\delta/5}}\,.
\eeq
If \,$\delta\,>\,1/3$\, then in \eq{ScSu} \,$f_{s}\rightarrow 1$\, at a time \,$\tau_{1}$\, which is given by
\beq \label{Rh17}
Q_{s}\tau_{1}\,\sim\,\al^{-15/(5 + 3\delta)}\,.
\eeq
In this case the scaling solutions give evolution much like the late stage of the ``bottom-up'' where the hard gluons
feed energy in the thermalized system of the soft gluons causing the temperature to rise with time until the full
thermalization of the system. The exception is that the gluons produced early at time $\tau_{0}$ now play the part of the
hard particles since \,$N_{s}(\tau,\tau_{0})\,>N_{h}$.\, Besides, \,$k_{s}(\tau_{0})$\, operates as the branching momentum
$k_{br}$ in \eq{Rh15}. As the time grows larger and larger the gluons from \,$N_{s}(\tau,\tau_{0})$\, at smaller
and smaller $\tau_{0}$ disappear into the thermal bath, through branching as in the ``bottom-up''. Equating the energy flow
from these gluons into the thermal bath of temperature $T$, the equation governing the evolution will be obtained:
\beq \label{Rh18}
\frac{d\epsilon}{d\tau}\,\sim\,T^{3}\,\frac{dT}{d\tau}\,\sim\,\frac{N_{s}(\tau,\tau_{0})}{\tau}\,k_{s}(\tau_{0})
\eeq
with
\beq \label{Rh19}
k_{s}(\tau_{0})\,\,\sim\,\al^{4}T^{3}\tau^{2}\,.
\eeq
Inserting \eq{Rh19} and the number density from \eq{Rh16} into \eq{Rh18} one obtains
\beq \label{Rh20}
T\,\sim\,Q_{s}\,\al^{\,\frac{35 - 78\delta}{39\delta - 10}}\,(Q_{s}\tau)^{\,\frac{15 - 36\delta}{39\delta - 10}}\,.
\eeq
The heating of the soft gluons thermal bath is finished when the energy transfer to the bath is complete. This takes
place when \,$k_{s}(\tau_{0})$\, reaches \,$Q_{s}$,\, namely from \eq{Rh19} and \eq{Rh20} we will have
\beq \label{Rh21}
Q_{s}\tau\,\sim\,\al^{-\frac{65 - 78\delta}{25 - 30\delta}}
\eeq
where at \,$\delta > 1/3$\, practically there is no dependence on the value of \,$\delta$\, and the resulting
\,$Q_{s}\tau\,\sim\,\al^{-13/5}$\, is the equilibration time of the original ``bottom-up'' thermalization.

Inserting \eq{Rh21} into \eq{Rh20} one will obtain the {\em m}``bottom-up'' temperature of the thermalized sector:
\beq \label{Rh22}
T\,\sim\,Q_{s}\,\al^{\theta(\delta)},\,\,\,\,\,\,\,\,\,\,\,\,\,\,\,\,\,\,\,\,
\theta(\delta)\,=\,\frac{(35 - 78\delta)(25 - 30\delta) - (65 - 78\delta)(15 - 36\delta)}{(39\delta - 10)(25 - 30\delta)}
\eeq
where at \,$\delta > 1/3$\, again the dependence on the value of \,$\delta$\, is negligible and \,$T\,\sim\,Q_{s}\,\al^{2/5}$.
We consider \eq{Rh21} and \eq{Rh22} as the equilibration time and temperature of the thermalized
gluon system in the {\em m}``bottom-up'' ansatz, however, coincided to those in the original ``bottom-up'' ansatz at
\,$1/3 \leq \delta \leq 10/21$\, \footnote{When \,$\delta = 10/21$,\, \,$N_{s}k_{s}$\, is equal to the energy density of
the hard gluons which is maximal energy carried by the soft gluons.}. Therefore, taking \,$\delta\,\geq\,1/3$\, we will not
distinguish the {\em m}``bottom-up'' from the original ``bottom-up''.

We are interested in difference of the gluons number between the late and initial stages (states), such that at RHIC and
higher energies in the central region of rapidity, \,$y\,\leq\,0$,\, the following ratio is defined
\cite{BMSS}:
\beq \label{Rh23}
R\,=\,[N_{s}(\tau)(Q_{s}\tau)]|_{\tau_{eq}}\,/\,\,[N_{h}(\tau)(Q_{s}\tau)]|_{\tau_{in}}\,\sim\,\al^{-2/5}\,>\,1\,
\eeq
which describes the branching process wherein the number of the gluons increases with \,$\tau$.
The ``bottom-up'' equilibration time and temperature are expressed via
\beq \label{Rh24}
\tau_{eq}\,=\,\varepsilon_{eq}\,\,\al^{-13/5}(Q_{s}^{2})\,Q_{s}^{-1}
\eeq
and
\beq \label{Rh25}
T_{eq}\,\simeq\,0.165\,\,\varepsilon\,\,\varepsilon_{eq}\,\,\al^{2/5}(Q_{s}^{2})\,Q_{s}
\eeq
where $\varepsilon_{eq}$ is the equilibration constant. Then by means of these two expressions the ratio $R$ defined in
\eq{Rh23} is found to be
\beq \label{Rh26}
R\,\simeq\,0.13\,\,\varepsilon^{2}\,\,\varepsilon_{eq}^{4}\,\al^{-2/5}(Q_{s}^{2})\,,
\eeq
and finally, for the most central collisions the result for the charged hadron multiplicity can be rewritten as
{\setlength\arraycolsep{2pt}
\bea \label{Rh27}
\left<\frac{2}{N_{part}}\frac{dN_{ch}}{dy} \right> & \simeq & 0.28\,R\,\varepsilon\,\ln\!\!\Lb
{\frac{\bar{Q}_{s}^{2}}{\Lambda_{QCD}^{2}}} \Rb
\nonumber\\
&\simeq & \,0.032\,\varepsilon^{3}\,\varepsilon_{eq}^{4}\left[\ln\!\!\Lb{\frac{\bar{Q}_{s}^{2}} {\Lambda_{QCD}^{2}}} \Rb\right]^{7/5}
\eea}
$\!\!\!$which substitutes for \eq{Rh14}. The Jacobian of the \,$y\leftrightarrow\eta$\, transformation at \,$y = \eta = 0$\,
is close to unity and gives the value $0.96$ for \,$Q_{s}\,=\,1\,GeV$\, taking the mass of the produced hadron
\,$m_{h}\,\simeq\,300\,MeV$\, \cite{Gel}.
In order to revisit the saturation constraints at RHIC (found in \cite{BMSS}) we equalise the charged hadron multiplicity in
\eq{Rh27} with the result of the PHOBOS Collaboration \cite{PHOBOS} at \,$W\,=\,130\,GeV$ for the most central collisions:
\beq \label{Rh28}
\left<\frac{2}{N_{part}}\frac{dN_{ch}}{d\eta} \right>\,=\,3.24\,.
\eeq
This experimental value corresponds to the theoretical one of \eq{Rh27} when
\beq \label{Rh29}
\varepsilon_{eq}\,\simeq\,\frac{2.13}{\varepsilon^{3/4}}
\eeq
which should be confronted with the consistency requirement of the ``bottom-up'' scenario, i.e., the ratio
$R$ in \eq{Rh26} must be larger than $2$:
\beq \label{Rh30}
\varepsilon^{2}\,\,\varepsilon_{eq}^{4}\,\geq\,11\,\,\Rightarrow\,\,\varepsilon_{eq}\,\geq\,\frac{1.82}{\sqrt{\varepsilon}}\,.
\eeq
\eq{Rh29} and \eq{Rh30} constrain two parameters:
\beq \label{Rh31}
\varepsilon\,\leq\,1.87\,\,\,\,\,\,\,\,\mbox{and}\,\,\,\,\,\,\,\,\varepsilon_{eq}\,\geq\,1.33\,.
\eeq
Another constraint which is due to formation of the equilibrated plasma (quark-gluon plasma), should
be also imposed additionally, namely
\beq \label{Rh32}
T_{eq}\,\geq\,T_{deconf}
\eeq
where $T_{deconf}$ is the phase transition temperature of the order of $180.5 MeV$ and $192.25 MeV$, respectively, for $3$
and $2$ flavour QCD found in current lattice studies \cite{Cheng}. These numbers are about $15^{\circ}\!\!/\!_{\circ}$
and $10^{\circ}\!\!/\!_{\circ}$ larger than the previously quoted values $154 \pm 8 MeV$ and $173 \mp 8 MeV$ \cite{KLP}.
Making use of \eq{Rh25} and \eq{Rh29} one obtains $\varepsilon\,>\,0.26$. Thus the parameters $\varepsilon$ and $\varepsilon_{eq}$
lie in the given limited range. In particular, at \,$\varepsilon \simeq 1.09$\, we have \,$\varepsilon_{eq} \simeq 2$ which
was used in \cite{BMSS}. At the end we wish to stress that in different approaches the saturation scale at RHIC varies in
the range of \,$1\,\div\,2\,GeV^{2}$.\, And it would be interesting to find the saturation constraints
for the maximum estimated $Q_{s}$ at RHIC. One can do that by taking \,$Q_{s}^{2}(b = 0)\,=\,2.05\,GeV^{2}$\, at
\,$W = 130\,GeV$\, \cite{KLN}. Then by means of \eq{Rh27}, \eq{Rh28} and \eq{Rh32} we receive
\beq \label{Rh33}
0.12\,\leq\,\varepsilon\,\leq\,1.5\,\,\,\,\,\,\,\,\mbox{and}\,\,\,\,\,\,\,\,\varepsilon_{eq}\,\geq\,1.46\,.
\eeq
Ultimately, in the kinematical range of RHIC, \,$Q_{s}^{2}\,\simeq\,2\,\div\,1\,GeV^{2}$,\, the algebraic average of the
gluon liberation coefficient varies in the range of \,$\varepsilon\,\simeq\,0.81\,\div\,1.06$\, respectively.

Now let us return to \eq{Sp15} with $m$ replaced by $m_{\infty}$:
\beq \label{Rh34}
\frac{m_{\infty}}{m_{0}}\,\approx\,1.35\left[2\,\al\,m_{0} \Lb \tau - \frac{\tau_{0}} {\gamma}\ln\!\!\Lb \frac{1 +
e^{\gamma(\tau/\tau_{0})}}{2} \Rb \Rb \right]^{1/4}\,\,\,\,\,\,\,\,\,\mbox{for $N_{c} = 3$}\,.
\eeq
Using \,$m_{\infty}\,=\,m_{D}/\sqrt{2}$\, along with $m_{D}$ from \eq{ScSu}
\beq \label{Rh35}
m_{D}\,=\,\frac{\beta\,Q_{s}}{(Q_{s}\tau)^{1/2 - 3\delta/10}}
\eeq
we will obtain
\beq \label{Rh36}
m_{0}(\al,\beta,\gamma,\delta,\tau)\,\approx\,0.6\,\frac{\beta^{4/5}}{\al^{1/5}}\,\frac{Q_{s}^{(10 +
6\delta)/25}}{\tau^{(10-6\delta)/25}}\,\frac{\gamma^{1/5}}{[2\gamma\tau - 2\tau_{0}\ln\!{((1 +
e^{\gamma(\tau/\tau_{0})})/2})]^{1/5}}\,.
\eeq
However, for numerical computations we need a constant value of $m_{0}$. Consequently, in order to do that, $\al$, $\beta$,
$\gamma$, $\delta$ and $\tau$ should be defined for the purpose of averaging of \,$m_{0}(\al,\beta,\gamma,\delta,\tau)$.\,
Note that the factor $\beta$ in \eq{Rh35} is introduced to convert the sign \,$\sim$\, into the sign \,$=$.

The factor \,$\beta$\, is unknown but realizing phenomenological comparisons of $m_{D}$, defined by \eq{Rh35}, with
calculations of a fixed and $T$-dependent Debye mass \cite{Zak,Kac} give $\beta \approx 1.8$ at RHIC and $\beta \approx 1.6$
at LHC. Of course, the value of $\beta$ should be the same for two cases but the uncertainty in our extraction is about
$11^{\circ}\!\!/\!_{\circ}$. The suggested range for $\gamma$ is [0, 6] which can be seen in \fig{rate_energy}. For the
number $\delta$ we take the interval \,$1/3<\delta<10/21$\, established for the late stage in the {\em m`}`bottom-up'' thermalization. Then for $\tau$ we take
the range \,$[0, \tau_{eq}]$\, making use of \eq{Rh24} with \,$\varepsilon_{eq}\,=\,2$.\, But for a parametrical comparison
with \,$\varepsilon_{eq}\,=\,2$\, case we will also perform calculations taking \,$\varepsilon_{eq}\,=\,1$,\, though it
is not supported by the saturation constraints (see \eq{Rh31}). QCD coupling is taken from \eq{Rh4} employing the value
\,$Q_{s}^{2}\,=\,1\,GeV^{2}$,\, being the saturation scale at RHIC. So that if we average $m_{0}$ over the ranges of
$\gamma$ (and at $\gamma\rightarrow0$), $\delta$ and $\tau$ for fixed $\al$ and $\beta$ then the value of $m_{0}$ at RHIC
can be defined as
\beq \label{Rh37}
m_{0}\,\equiv\,<m_{0}>\,=\,\frac{1}{\Delta\gamma\Delta\delta\Delta\tau}\int_{\gamma_{1}}^{\gamma_{2}}\!\!\!\!
\int_{\delta_{1}}^{\delta_{2}}\!\!\!\!\int_{\tau_{1}}^{\tau_{2}}m_{0}(\al,\beta,\gamma,\delta,\tau)\,d\gamma\,d\delta\,d\tau\,.
\eeq
Carrying out numerical estimates of this integral and confronting the results with \,$m_{0}\,\sim\,Q_{s}$ from \eq{Sp71}
(see \fig{trans1}) one can ascertain that at RHIC it is reasonable to take the lower limit of $m_{0}$ such as
\,$m_{0}\,\sim\,3\,\div\,3.5\,fm^{-1}$\, and the higher limit such as \,$m_{0}\,\sim\,5\,\div\,7\,fm^{-1}$.\,
Our procedure is somewhat rough but by this way we cover the possible range for $m_{0}$.

At this moment we can already match the time equilibration conditions of the {\em m}``bottom-up''/``bottom-up'' thermalization
and QCD wave turbulence onto each other, i.e.,
\beq \label{Rh38}
Q_{s}\tau_{eq}\,\approx\,\varepsilon_{eq}\,\al^{-13/5}\,\,\,\,\,\,\,\,\,\,\mbox{and}\,\,\,\,\,\,\,\,\,\,
m_{0}\Psi(\gamma,\tau_{eq}) \gg \al^{-9/5}\,.
\eeq

\begin{center}
\begin{footnotesize}
\TABLE[ht]{
\begin{tabular}{l|l|l|l|l}
\hline
$\,\,\,\gamma$ & $\,\,\,\,\,\,\,m_{0}$ & $\tau_{eq.starting}$ &
$ \,\,\,m_{0}\Psi(\gamma,\tau_{eq})/\al^{-9/5}$ & $[(\omega_{1} - \omega_{2}) + T/f_{q_{2}} - T/f_{q_{1}}](1/m)$ \\
& \,\,(fm$^{-1}$) & \,\,\,\,\,\,\,(fm) & \hskip 0.3truecm $ \varepsilon_{eq}=1\,\,|\,\varepsilon_{eq}=2$ & \hskip
1.55truecm $ \varepsilon_{eq}=1\,|\,\epsilon_{eq}=2$ \\
\hline %
\hline\
0.0 & \hskip 0.4truecm 4.0 & \hskip 0.5truecm 1.12 & \hskip 0.825truecm $1.54\,|\,3.08$ & \hskip 2.0truecm $0.58\,|\,0.25$ \\\

0.0 & \hskip 0.4truecm 5.0 & \hskip 0.5truecm 0.90 & \hskip 0.825truecm $1.92\,|\,3.85$ & \hskip 2.0truecm $0.44\,|\,0.19$ \\
\hline %
\hline\
0.5 & \hskip 0.4truecm 4.0 & \hskip 0.5truecm 1.02 & \hskip 0.825truecm $1.85\,|\,4.93$ & \hskip 2.0truecm $0.46\,|\,0.14$ \\\

0.5 & \hskip 0.4truecm 5.0 & \hskip 0.5truecm 0.83 & \hskip 0.825truecm $2.31\,|\,6.16$ & \hskip 2.0truecm $0.35\,|\,0.10$ \\
\hline %
\hline\
1.0 & \hskip 0.4truecm 4.0 & \hskip 0.5truecm 0.92 & \hskip 0.825truecm $2.46\,|\,11.1$ & \hskip 2.0truecm $0.32\,|\,0.50\cdot10^{-1}$ \\\

1.0 & \hskip 0.4truecm 5.0 & \hskip 0.5truecm 0.77 & \hskip 0.825truecm $3.08\,|\,13.8$ & \hskip 2.0truecm $0.25\,|\,0.37\cdot10^{-1}$ \\
\hline %
\hline\
2.0 & \hskip 0.4truecm 4.0 & \hskip 0.5truecm 0.76 & \hskip 0.825truecm $5.53\,|\,90.0$ & \hskip 2.0truecm $0.12\,|\,0.36\cdot10^{-2}$ \\\

2.0 & \hskip 0.4truecm 5.0 & \hskip 0.5truecm 0.65 & \hskip 0.825truecm $6.91\,|\,112$ & \hskip 1.1truecm $0.90\cdot10^{-1}\,|\,0.27\cdot10^{-2}$ \\
\hline %
\hline\
3.0 & \hskip 0.4truecm 4.0 & \hskip 0.5truecm 0.65 & \hskip 0.8truecm $15.0\,|\,957$ & \hskip 1.1truecm $0.30\cdot10^{-1}\,|\,0.19\cdot10^{-3}$ \\\

3.0 & \hskip 0.4truecm 5.0 & \hskip 0.5truecm 0.57 & \hskip 0.8truecm $18.7\,|\,1.20\cdot10^{3}$ & \hskip 1.1truecm $0.26\cdot10^{-1}\,|\,0.14\cdot10^{-3}$ \\
\hline %
\hline\
4.0 & \hskip 0.4truecm 4.0 & \hskip 0.5truecm 0.56 & \hskip 0.8truecm $45.0\,|\,1.14\cdot10^{4}$ & \hskip 1.1truecm $0.86\cdot10^{-2}\,|\,0.84\cdot10^{-5}$ \\\

4.0 & \hskip 0.4truecm 5.0 & \hskip 0.5truecm 0.50 & \hskip 0.8truecm $56.2\,|\,1.43\cdot10^{4}$ & \hskip 1.1truecm $0.65\cdot10^{-2}\,|\,0.64\cdot10^{-5}$ \\
\hline %
\hline\
5.0 & \hskip 0.4truecm 4.0 & \hskip 0.5truecm 0.50 & \hskip 0.875truecm $144\,|\,1.46\cdot10^{5}$ & \hskip 1.1truecm $0.20\cdot10^{-2}\,|\,0.35\cdot10^{-6}$ \\\

5.0 & \hskip 0.4truecm 5.0 & \hskip 0.5truecm 0.45 & \hskip 0.875truecm $180\,|\,1.83\cdot10^{5}$ & \hskip 1.1truecm $0.15\cdot10^{-2}\,|\,0.26\cdot10^{-6}$ \\
\hline %
\hline\
6.0 & \hskip 0.4truecm 4.0 & \hskip 0.5truecm 0.45 & \hskip 0.875truecm $479\,|\,1.94\cdot10^{6}$ & \hskip 1.1truecm $0.45\cdot10^{-3}\,|\,0.14\cdot10^{-7}$ \\\

6.0 & \hskip 0.4truecm 5.0 & \hskip 0.5truecm 0.41 & \hskip 0.875truecm $598\,|\,2.43\cdot10^{6}$ & \hskip 1.1truecm $0.34\cdot10^{-3}\,|\,0.10\cdot10^{-7}$ \\
\hline
\end{tabular}
\caption{The calculation of the values of QCD wave turbulence equilibration condition from \eq{Sp17} at different
values of $m_{0}$ and $\gamma$ using $\tau_{eq}$ ($\simeq 3.5\,fm$ at $\varepsilon_{eq}=2$) of the ``bottom-up''
thermalization from \eq{Rh24}. \,$\tau_{eq.starting}$\, is the time at which the equilibration in QCD wave turbulence
begins to set in, coming from the condition \,\,$m_{0}\Psi (\gamma,\tau_{eq.starting})\,=\,\al^{-9/5}$.\, The calculation
has been realized for \,$\al(Q_{s}^{2}\,=\,1\,GeV^{2})\,\approx\,0.43$.}}
\end{footnotesize}
\end{center}

In this connection one should point out that for this matching we rather need to consider the different values of $m_{0}$
instead of its fixed value, in order to see the behaviour of the matching along with $\gamma$'s.
In the Table 1 one can see the values of QCD wave turbulence equilibration condition using the following expression from
\eq{Sp17}:
\beq \label{Rh39}
\frac{1}{\al^{9/4}\left[m_{0}\Psi(\gamma,\tau_{eq})\right]^{5/4}}\,=\,[(\omega_{1}-\omega_{2}) + T/f_{q_{2}} -
T/f_{q_{1}}]\,\frac{1}{m}
\eeq
where $\tau_{eq}\,\equiv\,\tau_{eq}$(``bottom-up'',RHIC) for \,$\varepsilon_{eq}\,=\,1,2$\, from \eq{Rh24}. The
calculation have been fulfilled for two values of $m_{0}$ at $\gamma$'s by increasing order.
It is clear that at higher values of $m_{0}$ in each $\gamma$ column and/or at non-zero values of $\gamma$, the expression
\,$[(\omega_{1} - \omega_{2}) + T/f_{q_{2}} - T/f_{q_{1}}](1/m)$\, approaches to zero faster than at smaller values of
$m_{0}$ and at $\gamma = 0$, whereby QCD wave turbulence time equilibration picture becomes closer to that of the ``bottom-up''
thermalization, i.e., \,\,$\tau_{eq}(\mbox{QCD\,\,wave\,\,turbulence}) \rightarrow \tau_{eq}(\mbox{``bottom-up''}).$
For the matching the necessary requirement is \,$\tau_{eq.starting}<\tau_{eq}$(``bottom-up'')\, which is obtained by means
of the incoming gluon rate \,$\dot{\epsilon}(\tau)\,=\,\frac{m_{0}^{5}}{\al}\,\frac{2}{1 + e^{\gamma(\tau/\tau_{0})}}$.
We admit that for one of the values of the flow weakening parameter $\gamma$, it is much presumable that the coincidence
between the equilibration times of the gluon system in QCD turbulence scenario and ``bottom-up'' thermalization does occur.
Consequently, we suppose that there is a fixed value of $\gamma$ which with the genuine $m_{0}$ gives the matching
(coincidence) at RHIC and/or LHC (see Discussions and Conclusions).

\subsection{The picture at LHC}

We proceed our discussion to LHC energy \,$W\,=\,5500\,GeV$\, but before, one must define the saturation scale at LHC using
a simple formula from \cite{KLN} which for \,$y = 0$\, gives
\beq \label{Lh1}
Q_{s}^{2}(W)/Q_{s}^{2}(W_{0})\,=\,\Lb\frac{W}{W_{0}} \Rb^{\tilde{\lambda}}
\eeq
where \,$\tilde{\lambda}\,=\,\lambda/(1 + \frac{1}{2}\lambda)\,=\,0.252$\, with $\lambda\,=\,0.288$ \cite{GBW}.
We take \,$Q_{s}^{2}(W_{0}) = 2.05\,GeV^{2}$\, at \,$W_{0}\,=\,130\,GeV$\, whereby the interpolation formula gives
$Q_{s}^{2}(W)\,\simeq\,5.3\,GeV^{2}$ at \,$W\,=\,5500\,GeV$.\, Here $Q_{s}^{2}(W_{0}) = 1\,GeV^{2}$ is not applicable
since the resulting \,$Q_{s}(W)$\, does not satisfy the conventional condition for LHC energies, namely
\,$Q_{s}^{2}\,\geq\,4\,GeV^{2}$. It should be noted that the value \,$2.05\,GeV^{2}$\, is fixed at mid-rapidity and
\,$b = 0$\, for \,$N_{part}\,\simeq\,378$\, from the description of RHIC data on the multiplicity in $Au - Au$ collisions,
as computed in Glauber approach. Despite the small difference between atomic numbers of the gold and lead nuclei we make
use of \eq{Lh1} for getting the saturation scale in $Pb - Pb$ collisions at LHC \footnote{From numerical calculations
it is known that when normalized to the number of participants the multiplicity in the central \,$Au - Au$\, and
\,$Pb - Pb$\, collisional systems is almost identical, so that the extrapolated gold-$Q_{s}$ at \,$W\,=\,5500\,GeV$ can
be used instead of the lead-$Q_{s}$ at the same center-mass energy.}.

Now we need to modify \eq{Rh14} for LHC energy. In order to get the gluon structure function applicable for this case we
find (using the MRST parton distributions at NNLO \cite{TMS})
\beq \label{Lh2}
x G(x, Q_{s}^{2})\,=\,0.87\,\ln\!\!\Lb \frac{Q_{s}^{2}}{\Lambda_{QCD}^{2}} \Rb
\eeq
such that \,$x G(x, Q_{s}^{2})\,\simeq\,4.2$\, at \,$Q_{s}^{2}\,\simeq\,5\,GeV^{2}$\, (at \,$x\,\simeq\,8\cdot\,10^{-4}$\,
and \,$W\,=\,5500\,GeV$). Then the charged hadron multiplicity will be
\beq \label{Lh3}
\left<\frac{2}{N_{part}}\frac{dN_{ch}}{d\eta} \right>\,\simeq\,0.58\,\varepsilon\,\ln\!\!\Lb \frac{\bar{Q_{s}^{2}}}
{\Lambda_{QCD}^{2}} \Rb\
\eeq
and \eq{Rh27} can be rewritten as
{\setlength\arraycolsep{2pt}
\bea \label{Lh4}
\left<\frac{2}{N_{part}}\frac{dN_{ch}}{d\eta} \right> & \simeq & 0.58\,R\,\varepsilon\,\ln\!\!\Lb
{\frac{\bar{Q}_{s}^{2}}{\Lambda_{QCD}^{2}}} \Rb
\nonumber\\
&\simeq & \,0.066\,\varepsilon^{3}\,\varepsilon_{eq}^{4}\left[\ln\!\!\Lb{\frac{\bar{Q}_{s}^{2}} {\Lambda_{QCD}^{2}}}
\Rb\right]^{7/5}\,.
\eea}
$\!\!$For finding the saturation constraints at LHC we equalise the charged hadron multiplicity in \eq{Lh4} with
compilation of the PHOBOS results from \cite{PHOBOS1}
\beq \label{Lh5}
\left< \frac{2}{N_{part}}\frac{dN_{ch}}{d\eta}\right>\,=\,6.2
\eeq
which for \,$N_{part} = 386$\, corresponds to \,$dN_{ch}/d\eta\,=\,1200$. The value in \eq{Lh5} meets the theoretical
expectation of \eq{Lh4} when
\beq \label{Lh6}
\varepsilon_{eq}\,\simeq\,\frac{1.78}{\varepsilon^{3/4}}
\eeq
which again should be confronted with the consistency requirement of the ``bottom-up'' scenario:
\beq \label{Lh7}
\varepsilon^{2}\,\,\varepsilon_{eq}^{4}\,\geq\,9.4\,\,\Rightarrow\,\,\varepsilon_{eq}\,\geq\,\frac{1.75}{\sqrt{\varepsilon}}\,.
\eeq
\eq{Lh6} and \eq{Lh7} constrain $\varepsilon$ and $\varepsilon_{eq}$:
\beq \label{Lh8}
\varepsilon\,\leq\,1.09\,\,\,\,\,\,\,\,\mbox{and}\,\,\,\,\,\,\,\,\varepsilon_{eq}\,\geq\,1.67\,.
\eeq
Using \eq{Rh25} and \eq{Rh32} we find the lower limit as \,$\varepsilon\,>\,0.04$.\, As regards the condition
\,$\varepsilon_{eq}\,\simeq\,2$\, used for RHIC \cite{BMSS}, here it is received when \,$\varepsilon\,\simeq\,0.86$
\footnote{It must be stressed that numerical and analytical calculations of the gluon liberation coefficient yield results
which approximately vary in the range of $0.4\,-\,1.4$ \cite{Xi,Kras,Lap,Leo,Kov01}.}.
In addition to this picture one can realize an estimate of the saturation constraints using another conventionally
applicable value of the saturation momentum at LHC, \,$Q_{s}\,=\,3\,GeV$.\, Then by means of \eq{Lh4}, \eq{Lh5} and
\eq{Rh32} one obtains
\beq \label{Lh9}
0.02\,\leq\,\varepsilon\,\leq\,0.98\,\,\,\,\,\,\,\,\mbox{and}\,\,\,\,\,\,\,\,\varepsilon_{eq}\,\geq\,1.75\,.
\eeq
Thus at LHC when \,$Q_{s}^{2}\,\simeq\,9\,\div\,5.3\,GeV^{2}$,\, the algebraic average of the gluon
liberation coefficient varies in the range of \,$\varepsilon\,\simeq\,0.5\,\div\,0.56$\, respectively.

\begin{center}
\begin{footnotesize}
\TABLE[ht]{
\begin{tabular}{l|l|l|l|l}
\hline
$\,\,\,\gamma$ & $\,\,\,\,\,\,\,m_{0}$ & $\tau_{eq.starting}$ &
$ \,\,\,m_{0}\Psi(\gamma,\tau_{eq})/\al^{-9/5}$ & $[(\omega_{1} - \omega_{2}) + T/f_{q_{2}} - T/f_{q_{1}}](1/m)$ \\
& \,\,(fm$^{-1}$) & \,\,\,\,\,\,\,(fm) & \hskip 0.3truecm $ \varepsilon_{eq}=1\,\,|\,\varepsilon_{eq}=2$ & \hskip
1.6truecm $ \varepsilon_{eq}=1\,|\,\varepsilon_{eq}=2$ \\
\hline %
\hline\
0.0 & \hskip 0.4truecm 5.0 & \hskip 0.5truecm 1.90 & \hskip 0.8truecm $1.17\,|\,2.34$ & \hskip 2.0truecm $0.82\,|\,0.35$ \\\

0.0 & \hskip 0.4truecm 10.0 & \hskip 0.5truecm 0.95 & \hskip 0.8truecm $2.34\,|\,4.68$ & \hskip 2.0truecm $0.35\,|\,0.15$ \\
\hline %
\hline\
0.5 & \hskip 0.4truecm 5.0 & \hskip 0.5truecm 1.61 & \hskip 0.8truecm $1.51\,|\,4.59$ & \hskip 2.0truecm $0.60\,|\,0.15$ \\\

0.5 & \hskip 0.4truecm 10.0 & \hskip 0.5truecm 0.88 & \hskip 0.8truecm $3.02\,|\,9.18$ & \hskip 2.0truecm $0.25\,|\,0.63\cdot10^{-1}$ \\
\hline %
\hline\
1.0 & \hskip 0.4truecm 5.0 & \hskip 0.5truecm 1.36 & \hskip 0.8truecm $2.30\,|\,14.6$ & \hskip 2.0truecm $0.35\,|\,0.35\cdot10^{-1}$  \\\

1.0 & \hskip 0.4truecm 10.0 & \hskip 0.5truecm 0.80 & \hskip 0.8truecm $4.59\,|\,29.2$ & \hskip 2.0truecm $0.15\,|\,0.15\cdot10^{-1}$  \\
\hline %
\hline\
2.0 & \hskip 0.4truecm 5.0 & \hskip 0.5truecm 1.04 & \hskip 0.8truecm $7.31\,|\,259$ & \hskip 1.1truecm $0.83\cdot10^{-1}\,|\,0.96\cdot10^{-3}$ \\\

2.0 & \hskip 0.4truecm 10.0 & \hskip 0.5truecm 0.68 & \hskip 0.8truecm $14.6\,|\,518$ & \hskip 1.1truecm $0.35\cdot10^{-1}\,|\,0.40\cdot10^{-3}$ \\
\hline %
\hline\
3.0 & \hskip 0.4truecm 5.0 & \hskip 0.5truecm 0.84 & \hskip 0.8truecm $29.1\,|\,6.08\cdot10^{3}$ & \hskip 1.1truecm $0.15\cdot10^{-1}\,|\,0.19\cdot10^{-4}$ \\\

3.0 & \hskip 0.4truecm 10.0 & \hskip 0.5truecm 0.59 & \hskip 0.8truecm $58.2\,|\,1.22\cdot10^{4}$ & \hskip 1.1truecm $0.62\cdot10^{-2}\,|\,0.78\cdot10^{-5}$ \\
\hline %
\hline\
4.0 & \hskip 0.4truecm 5.0 & \hskip 0.5truecm 0.72 & \hskip 0.875truecm $130\,|\,1.60\cdot10^{5}$ & \hskip 1.1truecm $0.23\cdot10^{-2}\,|\,0.31\cdot10^{-6}$ \\\

4.0 & \hskip 0.4truecm 10.0 & \hskip 0.5truecm 0.52 & \hskip 0.875truecm $259\,|\,3.21\cdot10^{5}$ & \hskip 1.1truecm $0.96\cdot10^{-3}\,|\,0.13\cdot10^{-6}$ \\
\hline %
\hline\
5.0 & \hskip 0.4truecm 5.0 & \hskip 0.5truecm 0.63 & \hskip 0.875truecm $615\,|\,4.51\cdot10^{6}$ & \hskip 1.1truecm $0.33\cdot10^{-3}\,|\,0.48\cdot10^{-8}$ \\\

5.0 & \hskip 0.4truecm 10.0 & \hskip 0.5truecm 0.46 & \hskip 0.075truecm $1.23\cdot10^{3}\,|\,9.03\cdot10^{6}$ & \hskip 1.1truecm $0.14\cdot10^{-3}\,|\,0.20\cdot10^{-8}$ \\
\hline %
\hline\
6.0 & \hskip 0.4truecm 5.0 & \hskip 0.5truecm 0.56 & \hskip 0.075truecm $3.04\cdot10^{3}\,|\,1.32\cdot10^{8}$ & \hskip 1.1truecm $0.44\cdot10^{-4}\,|\,0.70\cdot10^{-10}$ \\\

6.0 & \hskip 0.4truecm 10.0 & \hskip 0.5truecm 0.42 & \hskip 0.075truecm $6.08\cdot10^{3}\,|\,2.65\cdot10^{8}$ & \hskip 1.1truecm $0.19\cdot10^{-4}\,|\,0.30\cdot10^{-10}$ \\
\hline
\end{tabular}
\caption{The calculation of the values of QCD wave turbulence equilibration condition from \eq{Sp17} at different
values of $m_{0}$ and $\gamma$ using $\tau_{eq}$ ($\simeq 4.4\,fm$ at $\varepsilon_{eq}=2$) of the ``bottom-up''
thermalization from \eq{Rh24}. \,$\tau_{eq.starting}$\, has the same meaning as in the Table 1. The calculation has been
realized for \,$\al(Q_{s}^{2}\,\approx\,5.27\,GeV^{2})\,\approx\,0.29$.}}
\end{footnotesize}
\end{center}

If we average $m_{0}$ over the ranges of $\gamma$ (and at $\gamma\rightarrow0$), $\delta$ and $\tau$ for fixed $\al$ and
$\beta$ then the numerical evaluations of the integral in \eq{Rh37} with the confrontation with \,$m_{0}\,\sim\,Q_{s}$
from \eq{Sp71} (see \fig{trans1}) gives that at LHC it is reasonable to take the lower limit of $m_{0}$ such as
\,$m_{0}\,\sim\,4\,\div\,4.5\,fm^{-1}$\, and the higher limit such as \,$m_{0}\,\sim\,11.5\,\div\,15\,fm^{-1}$\,. In the
Table 2 the values of QCD wave turbulence equilibration condition (from \eq{Sp17}) with \,$\epsilon_{eq}\,=\,1,2$\, at LHC
are shown for two values of $m_{0}$ at $\gamma$'s by increasing order. Again for the parametrical comparison with the case
\,$\varepsilon_{eq} = 2$\, we have carried out calculations taking \,$\varepsilon_{eq} = 1$\, which, nevertheless, is not
supported by the saturation constraints (see \eq{Lh8}).

\section{Discussions and Conclusions}
In this article we found the Kolmogorov gluon spectra in the presence of the low energy source which feeds in the energy
density at the time-dependent rate. The resulting equilibration time from the late stage spectrum was matched onto
$\tau_{eq}$ of the ``bottom-up'' thermalized system, however, the matching does depend on some selected parameters
which were discussed throughout this paper.

In Ref.\,\cite{MuShoWo06} there is also another considered case in which the incoming energy rate is spread uniformly in
the phase space in a region \,$0\,<\,k\,\leq\,k_{0}$\, of momenta. $k_{0}$ is a separate dimensional parameter to be a
scale larger than and independent of $m_{0}$, \,$k_{0}/m_{0}\,\geq\,1$,\, since the situation where $k_{0}$ is less than
$m_{0}$ seems to have an abnormally high rate of deposition of the energy over a limited region of the phase space. It
was argued that the complete thermalization occurs at a time
\beq \label{Con1}
m_{0}\tau\,\sim\,\al^{-9/5}\,\,\,\,\,\,\,\,\,\,\mbox{if}\,\,\,\,\,k_{0}/m\,\sim\,1
\eeq
and
\beq \label{Con2}
m_{0}\tau\,\sim\,\al^{-9/5}\Lb m/k_{0} \Rb^{12/5}\,\,\,\,\,\,\,\,\,\,\mbox{if}\,\,\,\,\,k_{0}/m\,\ll\,1\,.
\eeq
However, in our paper $k_{0}$ can be less than $m_{0}$ as well, because the abnormally high rate of deposition of the energy
over the limited region of the phase space is diminished based on use of the time-dependent rate in \eq{Sp1}. In any case
we did not derive the gluon spectra with the parameter $k_{0}$ since for numerical computations we would fix its value
arbitrarily. But we conjecture that \eq{Con1} and \eq{Con2} can also be generalized to our case of non-zero $\gamma$'s,
like the results of Sec 3.
Therefore, at genuine fixed $m_{0}$ (and $k_{0}$) the thermal equilibration time of QCD wave turbulence can be matched onto
(or coincide to) \,$\tau_{eq}$\, of the ``bottom-up'' thermalization depending on the energy flow weakening parameter
$\gamma$.

If we take \,$\tau_{eq}$\, of different evolutional scenarios after $AA$ collisions such as
\begin{itemize}
\item[1.] a) Hawking-Unruh radiation via the gluon emission off rapidly decelerating nuclei and b) multiparticle production
in the framework of the color glass condensate approach to high density QCD \cite{Kharz} where
\,$\tau_{therm}\,\simeq\,2\sqrt{2\pi}\,Q_{s}^{-1}\,\simeq\,1\,fm;\,0.4\,fm$,\, respectively, for
\,$Q_{s}\,=\,1\,GeV;\,2.3\,GeV$\,\,,

\item[2.] thermalization within microscopical parton cascade BAMPS (which is a microscopical transport model) \cite{El}
\footnote{In this approach, in agreement with the ``bottom-up'' thermalization, the equilibration time proves to be
proportional to \,$Q_{s}^{-1}$,\, nevertheless, its proportionality to \,$\al^{-13/5}$\, is not seen, but is much weaker
like \,$(\al\,\ln\!{(\al)})^{-2}Q_{s}^{-1}$.\, On the other hand, the thermal equilibrium of the soft and hard gluons
occurs roughly on the same time scale (due to \,$2 \leftrightarrow 3$\, processes) which contradicts the ``bottom-up''
picture while the energy flows into both soft and hard sectors at the same time which is potentially similar to the
phenomenon of the ``avalanche''.} where \,$\tau_{therm} = \al^{-2}(\ln{\al})^{-2}Q_{s}^{-1} \simeq
1.5\,fm;\,0.7\,fm$,\, respectively, for \,$\al\,\simeq\,0.43$,\, $Q_{s}\,=\,1\,GeV$\, and \,$\al\,\simeq\,0.29$,\,
$Q_{s}\,=\,2.3\,GeV$\,\,,

\item[3.] hydrodynamical evolution towards the equilibration based on corresponding models \cite{Kolb,Tean} where
\,$\tau_{eq}\,\leq\,0.5\,fm$
\end{itemize}
then the matching (or coincidence) with $\tau_{eq}$ of QCD wave turbulence can occur as well, only here $\gamma$ will take
higher values to be adequate for smaller equilibration times of these approaches. So that it is feasible to construct the
analogous Table 1 and 2 for the above and other evolutional approaches separately.

Notice that in our calculations there was small discrepancy in definition of QCD coupling $\al$. Matter of fact in Sec 4
we did the matchings using $\al$ defined in flavour QCD while in QCD wave turbulence one considers a purely gluonic system
with small $\al$. In addition, the equilibration time obtained from \eq{Rh24}, with $\al$ from \eq{Rh4}, is much larger at
RHIC and LHC compared to results of the hydrodynamical models \cite{Kolb,Tean}. It is due to admission \cite{BMSS} of
$\al$ dependence on the saturation scale of the colliding nuclei which is different at RHIC and LHC. Otherwise, if we used
the same value of $\al$, $\tau_{eq}$ would be smaller at LHC in contrast to RHIC (however, in the remaining part of this
section we realize phenomenological calculations with this ansatz).

In general, if one applied very small values of $\al$ (than had been used in this work), the gluon system would reach the
thermal equilibrium in much later times at the rate \,$\dot{\epsilon}\,=\, m_{0}^{5}/\al$\, which corresponds to
\,$\gamma = 0$. Consequently, in order to reduce $\tau_{eq|therm}$ of the system we did an assumption about time
dependence of the incoming gluon rate, introducing the parameter $\gamma$ which lowers the energy flow from the soft to hard
scales. The weakening parameter allows to decrease the evolutional time domain of the gluon system towards the thermal
equilibrium. Hereby, for fixed $m_{0}$ (and $k_{0}$ in general) and very small $\al$ one can choose the parameter $\gamma$
such that the derivable early, intermediate and late time Kolmogorov gluon spectra can be placed within time interval
\,$m_{0}^{-1}\,<\,\tau\,<\,\tau_{eq|therm}$\, where \,$\tau_{eq|therm}$\, is taken from various evolutional approaches
\footnote{However, the parameter $\gamma$ can be constrained and/or somewhat fixed. See discussion on the next page.}.
We consider the deduced ``running'' with time analytic gluon spectra as the main result of our paper but they, with the
performed numerical calculations, are mostly parametric.

Eventually, let us exhibit the values of the thermal equilibration time (see Table 3) computed from
\,$m_{0}\Psi(\gamma,\tau_{therm}) \sim \al^{-9/5}$\, if suppose that this is a modified form of \eq{Con1} at
\,$k_{0}/m \sim 1$,\, and this modification we take based on the discussion of Sec 3. But now contrary to Sec 3 and 4
will not consider the dependence of $\al$ on $Q_{s}$, instead in below discussion both for RHIC and LHC we will deal with
the values of \,$\tau_{therm}$'s\, at the same $\al$'s. In the Table 3 we show the values of \,$\tau_{therm}$\,
for \,$m_{0} = 1;\,5;\,10\,fm^{-1}$\, (albeit first is less than the lower limit obtained in Sec 4) at
\,$\al = 0.004;\,0.04;\,0.4$.\, Note that in fact the formulae of Sec 3, especially, \eq{Sp4} and \eq{Sp8}, do work much
better for very small values of $\al$ such as \,$\al\,=\,0.004$\, \cite{Leo}. Such very small \,$\al$'s\, are not applicable
in the ``bottom-up'' thermalization since the thermal equilibration time there takes much larger values.

\begin{center}
\begin{footnotesize}
\TABLE{
\begin{tabular}{l|l|l|l|l|l|l|l|l}
\hline
$\,\,$ & $\gamma\,=\,0.0$ & $\gamma\,=\,0.5$ & $\gamma\,=\,1.0$ & $\gamma\,=\,2.0$ & $\gamma\,=\,3.0$ &
$\gamma\,=\,4.0$ & $\gamma\,=\,5.0$ & $\gamma\,=\,6.0$ \\
\hline %
\hline\
$\al\,=\,0.004;\,m_{0}\,=\,1\,fm^{-1}$ & $20715\,fm$ & $23.7\,fm$ & $12.7\,fm$ & $6.79\,fm$ & $4.69\,fm$ & $3.61\,fm$ & $2.94\,fm$ & $2.49\,fm$  \\
\hline %
\hline\
$\al\,=\,0.004;\,m_{0}\,=\,5\,fm^{-1}$ & $4143\,fm$ & $19.7\,fm$  & $10.7\,fm$ & $5.78\,fm$ & $4.02\,fm$ & $3.11\,fm$ & $2.54\,fm$ & $2.16\,fm$  \\
\hline %
\hline\
$\al\,=\,0.004;\,m_{0}\,=\,10\,fm^{-1}$ & $2072\,fm$ & $17.9\,fm$ & $9.83\,fm$ & $5.35\,fm$ & $3.73\,fm$ & $2.89\,fm$ & $2.37\,fm$ & $2.01\,fm$  \\
\hline %
\hline\
$\al\,=\,0.04;\,m_{0}\,=\,1\,fm^{-1}$ & $328\,fm$ & $13.3\,fm$ & $7.53\,fm$ & $4.20\,fm$ & $2.97\,fm$ & $2.32\,fm$ & $1.91\,fm$ & $1.63\,fm$  \\
\hline %
\hline\
$\al\,=\,0.04;\,m_{0}\,=\,5\,fm^{-1}$ & $65.7\,fm$ & $9.31\,fm$ & $5.52\,fm$ & $3.19\,fm$ & $2.30\,fm$ & $1.81\,fm$ & $1.51\,fm$ & $1.29\,fm$  \\
\hline %
\hline\
$\al\,=\,0.04;\,m_{0}\,=\,10\,fm^{-1}$ & $32.8\,fm$ & $7.60\,fm$ & $4.66\,fm$ & $2.76\,fm$ & $2.01\,fm$ & $1.60\,fm$ & $1.33\,fm$ & $1.15\,fm$  \\
\hline %
\hline\
$\al\,=\,0.4;\,m_{0}\,=\,1\,fm^{-1}$ & $5.21\,fm$ & $3.33\,fm$ & $2.42\,fm$ & $1.62\,fm$ & $1.25\,fm$ & $1.02\,fm$ & $0.87\,fm$ & $0.77\,fm$  \\
\hline %
\hline\
$\al\,=\,0.4;\,m_{0}\,=\,5\,fm^{-1}$ & $1.04\,fm$ & $0.96\,fm$ & $0.87\,fm$ & $0.72\,fm$ & $0.62\,fm$ & $0.54\,fm$ & $0.48\,fm$ & $0.44\,fm$  \\
\hline %
\hline\
$\al\,=\,0.4;\,m_{0}\,=\,10\,fm^{-1}$ & $0.52\,fm$ & $0.50\,fm$ & $0.48\,fm$ & $0.43\,fm$ & $0.39\,fm$ & $0.36\,fm$ & $0.33\,fm$ & $0.31\,fm$  \\
\hline %
\end{tabular}
\caption{The thermal equilibration time from \,$m_{0}\Psi (\gamma,\tau_{therm})\,\sim\,\al^{-9/5}$\, at very small and
realistic \,$\al$'s\, when \,$k_{0}/m\,\sim\,1$.}}
\end{footnotesize}
\end{center}

In the Table 1,\,2 we considered $\gamma$ as an arbitrary parameter which allows us match \,$\tau_{therm}$\, to that of
the ``bottom-up'' thermalization. However, we suppose that the factor $\gamma$ can be fixed in a way. Dropping a look on
the Table 3 we see that when \,$\gamma \gtrsim 4$\, the difference of the highest and smallest \,$\tau_{therm}$'s\,
alters over one order of magnitude, in contrast to the case of \,$\gamma = 0$\, where the change is of the order of $10^4$.\,
Such a picture also approximately occurs when one performs calculations with \,$m_{0} = 3\,fm^{-1}$\, as the lower-bound at
RHIC and \,$m_{0} = 15\,fm^{-1}$\, as the upper-bound at LHC, both being defined in Sec 4. It is possible to check
that at the range of $\al$'s shown in the Table 3, the upper-bound and lower-bound of \,$\tau_{therm}$\, are such that
\beq \label{Con3}
\frac{\tau_{therm}(\mbox{upper-bound})}{\tau_{therm}(\mbox{lower-bound})}\,\approx\,7(8)\,\div\,10(12) \,\,\,\,\,\,\,\,\,
\mbox{at}\,\,\,\,\gamma\,\approx\,8\,\div\,4\,.
\eeq
On the one hand, at very small couplings and \,$\gamma \approx 6 \pm 2$\, our approach yields such values of \,$\tau_{therm}$\,
which are comparable to those from dynamical approaches (such as the ``bottom-up'' scenario) and on the other hand, at
realistic couplings and \,$\gamma \approx 6 \pm 2$\, it yields values of \,$\tau_{therm}$\, comparable to those from the
hydrodynamical models. So that it may be stated the following: the modified QCD wave turbulence at very small (even if
\,$\al\,=\,0.4\cdot 10^{-3}\,\,\,\,\mbox{or}\,\,\,\,0.4\cdot 10^{-4}$\,) and realistic couplings gives upper and lower
values of the thermal equilibration time to be of the same order as obtained, correspondingly, from perturbative (upper
\,$\tau_{therm}$)\, and hydrodynamical (lower \,$\tau_{therm}$)\, thermalization scenarios. Schematically depicted we have
the following approximate picture
\begin{displaymath}
\tau_{therm}(1)\,\lesssim\,\tau_{therm}(2)\,\lesssim\,\tau_{therm}(3)
\end{displaymath}
where
\begin{enumerate}
\item $\tau_{therm}(1)$ comes from the hydrodynamical models and/or multiparticle production in the framework of the color glass
condensate approach to high density QCD\,;
\item $\tau_{therm}(2)$ comes from the modified \,$(<\!\gamma\!>\,\approx\,6\,\pm\,2)$\, Kolmogorov wave turbulence in QCD\,;
\item $\tau_{therm}(3)$ comes from the dynamical approaches which are applied to RHIC and LHC energies\,.
\end{enumerate}
Thus at our chosen ranges of $\al$ and $m_{0}$ we roughly fix \,$<\!\gamma\!>\,\approx\,6$\, as a theoretically established
magnitude of the energy flow weakening parameter (but which concerns the modified case of \eq{Con1}). In this connection,
taking the mean values

I. \,\,\,\,$<\!m_{0}\!>\,\approx\,4.6\,fm^{-1}|0.91\,GeV$\, (from \,$m_{0}\,=\,3\,fm^{-1} - 3.5\,fm^{-1}\,\div\,5\,fm^{-1}
- 7\,fm^{-1}$\, \,\,at RHIC)\,;

II. \,$<\!m_{0}\!>\,\approx\,8.8\,fm^{-1}|1.73\,GeV$\, (from \,$m_{0}\,=\,4\,fm^{-1} - 4.5\,fm^{-1}\,\div\,11.5\,fm^{-1}
- 15\,fm^{-1}$\, \,\,at LHC)
then averaging the sum of all values of \,$\tau_{therm}$\, defined at realistic couplings  \,$\al = 0.1;\,0.105;...;\,0.4$,\,
and at small couplings \,$\al = 0.004;\,0.08;\,...;\,0.1$,\, we obtain
(after a designation \,$<\!\tau_{therm}\!>\,\equiv\,\tau_{therm}$\,)
\beq \label{Con4}
0.65\,fm\,\leq\,\tau_{therm}\,\leq\,1.30\,fm\, \,\,\,\,\mbox{at RHIC}\,\, \,\,\,\,\,\,\,\,\mbox{and}\,\,\,\,\,\,\,\,
\,0.52\,fm\,\leq\,\tau_{therm}\,\leq\,1.17\,fm\, \,\,\,\,\mbox{at LHC\,.}
\eeq
Notice that these phenomenological calculations are due to the selected interval of the coupling ranging in
$0.004\,\leq\,\al\,\leq\,0.4$. Nonetheless, if we took into account the joint modified picture of \eq{Con1} and \eq{Con2}
and even smaller $\al$'s instead of $0.004$, then the estimated lower and, especially, upper-bounds of \,$\tau_{therm}$\,
could be somewhat higher than given above which, in turn, would result to shifting of $\gamma$ from the roughly fixed
\,$\gamma \approx 6$. Consequently, we use this value as an initial one, emphasizing that determination of the exact
magnitude of the energy flow weakening parameter is a question of further investigations.

However, keeping the above line, we wish to present other phenomenological estimates of \,$\tau_{therm}$\, done at
realistic couplings (see Tables $4$ and $5$). This time we carry out calculations at \,$\gamma = 0$\, and \,$\gamma = 6$\,
which reflect the pictures of the original Kolmogorov wave turbulence in QCD and its modified version. We take into account
the intermediate regime of \,$\gamma = 3$\, as well, considering it as a middle case between the regime where the energy
amount deposited in the soft gauge sector is a constant in time \,($\gamma = 0$)\, and the regime where the energy amount
decreases exponentially but strongly with the above roughly fixed number \,$\gamma = 6$.

\begin{center}
\begin{footnotesize}
\TABLE{
\begin{tabular}{l|l|l|l}
\hline
$\,\,\,\,\,\,\,\,\,\,\,\,\,\,\,\,\,\,\,\,\,\,\,\,\,\,\,\,\,\,\,\,\,\,\,\,\,\,\,\,
\mbox{RHIC}$ & $\,\,\,\,\,\,\,\,\,\,\,\,\,\,\,\,\gamma\,=\,0.0$ & $\,\,\,\,\,\,\,\,\,\,\,\,\,\,\,\,\gamma\,=\,3.0$
& $\,\,\,\,\,\,\,\,\,\,\,\,\,\,\,\,\gamma\,=\,6.0$ \\
\hline %
\hline\
$\al\,=\,0.25;\,m_{0}\,=\,3\,fm^{-1}\,-\,7\,fm^{-1}$ & $4.04\,fm\,-\,1.73\,fm$ & $1.14\,fm\,-\,0.81\,fm$ & $0.71\,fm\,-\,0.54\,fm$ \\
\hline %
\hline\
$\al\,=\,0.30;\,m_{0}\,=\,3\,fm^{-1}\,-\,7\,fm^{-1}$ & $2.91\,fm\,-\,1.25\,fm$ & $1.01\,fm\,-\,0.68\,fm$ & $0.65\,fm\,-\,0.48\,fm$ \\
\hline %
\hline\
$\al\,=\,0.35;\,m_{0}\,=\,3\,fm^{-1}\,-\,7\,fm^{-1}$ & $2.21\,fm\,-\,0.95\,fm$ & $0.90\,fm\,-\,0.59\,fm$ & $0.59\,fm\,-\,0.42\,fm$ \\
\hline %
\hline\
$\al\,=\,0.40;\,m_{0}\,=\,3\,fm^{-1}\,-\,7\,fm^{-1}$ & $1.73\,fm\,-\,0.74\,fm$ & $0.81\,fm\,-\,0.50\,fm$ & $0.54\,fm\,-\,0.37\,fm$ \\
\hline %
\end{tabular}
\caption{$\tau_{therm}$\, at RHIC realistic \,$\al$'s\, when \,$k_{0}/m\,\sim\,1$.}}
\end{footnotesize}
\end{center}

\begin{center}
\begin{footnotesize}
\TABLE{
\begin{tabular}{l|l|l|l}
\hline
$\,\,\,\,\,\,\,\,\,\,\,\,\,\,\,\,\,\,\,\,\,\,\,\,\,\,\,\,\,\,\,\,\,\,\,\,\,\,\,\,
\mbox{LHC}$ & $\,\,\,\,\,\,\,\,\,\,\,\,\,\,\,\,\gamma\,=\,0.0$ & $\,\,\,\,\,\,\,\,\,\,\,\,\,\,\,\,\gamma\,=\,3.0$
& $\,\,\,\,\,\,\,\,\,\,\,\,\,\,\,\,\gamma\,=\,6.0$ \\
\hline %
\hline\
$\al\,=\,0.25;\,m_{0}\,=\,4\,fm^{-1}\,-\,15\,fm^{-1}$ & $3.03\,fm\,-\,0.81\,fm$ & $1.03\,fm\,-\,0.53\,fm$ & $0.65\,fm\,-\,0.39\,fm$ \\
\hline %
\hline\
$\al\,=\,0.30;\,m_{0}\,=\,4\,fm^{-1}\,-\,15\,fm^{-1}$ & $2.18\,fm\,-\,0.58\,fm$ & $0.90\,fm\,-\,0.43\,fm$ & $0.59\,fm\,-\,0.33\,fm$ \\
\hline %
\hline\
$\al\,=\,0.35;\,m_{0}\,=\,4\,fm^{-1}\,-\,15\,fm^{-1}$ & $1.65\,fm\,-\,0.44\,fm$ & $0.79\,fm\,-\,0.35\,fm$ & $0.53\,fm\,-\,0.28\,fm$ \\
\hline %
\hline\
$\al\,=\,0.40;\,m_{0}\,=\,4\,fm^{-1}\,-\,15\,fm^{-1}$ & $1.30\,fm\,-\,0.35\,fm$ & $0.70\,fm\,-\,0.29\,fm$ & $0.48\,fm\,-\,0.24\,fm$ \\
\hline %
\end{tabular}
\caption{$\tau_{therm}$\, at LHC realistic \,$\al$'s\, when \,$k_{0}/m\,\sim\,1$.}}
\end{footnotesize}
\end{center}

From the Tables 4 and 5 we see the upper and lower bounds of \,$<\!\tau_{therm}\!>$\, at \,$\gamma\,=\,0,\,3,\,6$.\,
Finally, realizing the above averaging procedure and doing the designation \,$<\!\tau_{therm}\!>\,\equiv\,\tau_{therm}$\,
our estimates can be represented as follows:
\beq \label{Con5}
0.45\,fm - 0.65\,fm\,\leq\,\tau_{therm}\,\leq\,0.97\,fm - 2.72\,fm\, \,\,\,\,\mbox{at RHIC}
\eeq
and
\beq \label{Con6}
0.31\,fm - 0.40\,fm\,\leq\,\tau_{therm}\,\leq\,0.86\,fm - 2.04\,fm\, \,\,\,\,\mbox{at LHC\,.}
\eeq

\vskip 0.0truecm %
\hskip -0.7truecm %
More precisely, we could take \,$\gamma \simeq 3.2$\, for the intermediate boundary regime since at this point
$$\frac{\mbox{energy\,\,flow\,\,rate}(\gamma\,=\,0)}{\mbox{energy\,\,flow\,\,rate}(\gamma\,=\,3.2)}\,\approx\,
\frac{\mbox{energy\,\,flow\,\,rate}(\gamma\,=\,3.2)}{\mbox{energy\,\,flow\,\,rate}(\gamma\,=\,6)}
$$
but the first upper-bounds of \,$\tau_{therm}$\, and its second lower-bounds would be slightly different from $0.97,0.86$
and $0.65,0.40$.

As stated before the Table 3, \eq{Sp4} and \eq{Sp8}, which make a contribution to the derivation of the gluon spectra, do
function well at very small couplings rather than at realistic $\al$'s. But in all our five tables we fully or partly used
the realistic couplings. Therefore, it is necessary to discuss the relevance of usage of the realistic coupling in our
paper. Let us take the values of $\al$ such as
$$
\al_{10} = 3\cdot 10^{-10};\,\al_{9} = 3\cdot 10^{-9};\,\al_{8} = 3\cdot 10^{-8};\,\al_{7} = 3\cdot 10^{-7};\,
\al_{6} = 3\cdot 10^{-6};\,\al_{5} = 3\cdot 10^{-5};
$$
$$
\al_{4} = 3\cdot 10^{-4};\,\al_{3} = 3\cdot 10^{-3};\,\al_{2} = 3\cdot 10^{-2};\,
\al_{1} = 0.3\,\,.
$$
If the gluon spectra in our paper would have not been derived doing the approximation of \eq{Sp9}, then at very small
$\al$'s new gluon spectra could yield such results for the proper time $\tau$ which would be identical to the results
appearing from our approximated spectra (notably, from the late stage spectrum in \eq{Sp17}).
In short, the quotients of the following ratios at any fixed $m_{0}$ and $\gamma$
$$
\frac{\tau_{therm}(\al_{10})}{\tau_{therm}(\al_{9})}\,\equiv\,a;\,\,\,
\frac{\tau_{therm}(\al_{9})}{\tau_{therm}(\al_{8})}\,\equiv\,b;\,\,\,
\frac{\tau_{therm}(\al_{8})}{\tau_{therm}(\al_{7})}\,\equiv\,c;\,\,\,
\frac{\tau_{therm}(\al_{7})}{\tau_{therm}(\al_{6})}\,\equiv\,d;
$$
$$
\frac{\tau_{therm}(\al_{6})}{\tau_{therm}(\al_{5})}\,\equiv\,e;\,\,\,
\frac{\tau_{therm}(\al_{5})}{\tau_{therm}(\al_{4})}\,\equiv\,f;\,\,\,
\frac{\tau_{therm}(\al_{4})}{\tau_{therm}(\al_{3})}\,\equiv\,g
$$
would have the same values both from our approximated late stage spectrum (see \eq{Sp17}) and from an exact late stage
spectrum (which is not derived in our paper). In any case at any fixed $m_{0}$ and $\gamma$ one can extrapolate the
chain \,$a,\,b,\,c,\,d,\,e,\,f,\,g$\, to the regime of the realistic couplings, expressed by
$$
\frac{\tau_{therm}(\al_{3})}{\tau_{therm}(\al_{2})}\,\equiv\,h;\,\,\,
\frac{\tau_{therm}(\al_{2})}{\tau_{therm}(\al_{1})}\,\equiv\,i
$$
and accomplishing sequential corrections at $\al_{2}$ and $\al_{1}$ we can find an approximately corrected
\,$\tau_{therm}(\al_{1})$\, without having the exact spectrum. Ultimately, it turns out that at extrapolated realistic
couplings the corrected values of the equilibration time do not differ noticeably from those obtained from \eq{Sp18}, and
it turns out that for the intermediate and higher $\gamma$'s, the change of \,$\tau_{therm}$\, is small as compared to the
change at \,$\gamma = 0$\, case. For example, in case of the exact derivation of \eq{Sp17} which could be applicable for all
$\al$'s, the lower-bounds of \,$\tau_{therm}$\, in \eq{Con4} would be higher but the difference with the upper-bounds
again would be noticeable.

So that under this circumstance we do final phenomenological estimates of \,$\tau_{therm}$\, at \,$<\!\gamma\!>\,=\,6$\, and
\,$<\!m_{0}\!>\,=\,4.6\,fm^{-1}$\, at RHIC and \,$<\!m_{0}\!>\,=\,8.8\,fm^{-1}$\, at LHC. Once again we take
\,$<\!\tau_{therm}\!>\,\equiv\,\tau_{therm}$\, which is the algebraical averaged value of the thermal equilibration time
calculated at the realistic couplings, \,$\al\,=\,0.25;\,0.3;\,0.35;\,0.4$\, (as in the Tables 4 and 5):
\beq \label{Con7}
0.53\,<\,\tau_{therm}\,<\,0.7\,fm\,\,\,\,\,\,\,\mbox{at\, RHIC\,\,\,\,\,\,\,\,\,\,\,\,\,\,\,\,\,\,\,and}
\,\,\,\,\,\,\,\,\,\,\,\,\,\,\,\,\,\,\,0.41\,<\,\tau_{therm}\,<\,0.65\,fm\,\,\,\,\,\,\,\mbox{at\, LHC}
\eeq
where the lower-bounds directly come from \eq{Sp18}, and the upper-bounds from the extrapolation of
\,$a,\,b,\,c,\,d,\,e,\,f,\,g$\, but again making use of \eq{Sp18}. Increasing accuracy of this extrapolation shows that
more precise values of \,\,$\tau_{therm}(realistic\,\,couplings)$\,\, lie in these ranges.
Thus summarizing our discussions in this section we arrive at \,$\tau_{therm}$'s\, obtained from
\begin{itemize}
\item[a)] combination of the original and modified QCD wave turbulent scenarios (see \eq{Con5} and \eq{Con6});
\item[b)] combination of very small and realistic couplings in the modified QCD wave turbulent scenario (see \eq{Con4});
\item[c)] combination of the realistic couplings and extrapolated realistic couplings in the modified QCD wave turbulent
scenario (see \eq{Con7}).
\end{itemize}
And we take also into consideration that if all numbers in \eq{Con4} and \eq{Con7} can be represented with satisfactory
accuracy, then the second upper-bounds of \eq{Con5} and \eq{Con6} at extrapolated realistic couplings (when $\gamma = 0$)
can change noticeably.

\section*{Acknowledgments}
I am grateful to Genya Levin for his careful reading of this manuscript and for very inspiring suggestions. This research
was supported in part by the Israel Science Foundation, founded by the Israeli Academy of Science and Humanities.

\end{document}